\documentclass[12pt]{article}
\usepackage{putex}
\usepackage{graphicx}
\def \ie {{\it i.e.}}

\def \half {{1 \over 2}}
\def \dalpha {{\Delta_S}}

\def \dbeta {{\Delta_\sigma}}

\def\NLsM{NL$\sigma$M}

\begin{document}

\preprint{hep-th/0512355 \\ PUPT-2189}

\institution{PU}{Joseph Henry Laboratories, Princeton University, Princeton, NJ 08544}

\title{Non-linear sigma models with anti-de Sitter target spaces}

\authors{Joshua J. Friess and Steven S. Gubser}

\abstract{We present evidence that there is a non-trivial fixed point for the $AdS_{D+1}$ non-linear sigma model in two dimensions, without any matter fields or additional couplings beyond the standard quadratic action subject to a quadratic constraint.  A zero of the beta function, both in the bosonic and supersymmetric cases, appears to arise from competition between one-loop and higher loop effects.  A string vacuum based on such a fixed point would have string scale curvature.  The evidence presented is based on fixed-order calculations carried to four loops (corresponding to $O(\alpha'^3)$ in the spacetime effective action) and on large $D$ calculations carried to $O(D^{-2})$ (but to all orders in $\alpha'$).  We discuss ways in which the evidence might be misleading, and we discuss some features of the putative fixed point, including the central charge and an operator of negative dimension.  We speculate that an approximately $AdS_5$ version of this construction may provide a holographic dual for pure Yang-Mills theory, and that quotients of an $AdS_3$ version might stand in for Calabi-Yau manifolds in compactifications to four dimensions.}

\date{December 2005}

\maketitle

\tableofcontents
\newpage

\section{Introduction}
\label{INTRODUCTION}

Consider the bosonic non-linear sigma model (\NLsM) in $d$
dimensions, whose target space is Euclidean anti-de Sitter space
($AdS_{D+1}$) with $D+1$ dimensions.  Explicitly, the classical
action is
 \eqn{SAdS}{
  S = {1 \over 4\pi\alpha'} \int d^dx \, (\partial n_\mu)^2
 }
where $n_\mu$ is constrained to satisfy
 \eqn{DefinAdS}{
  n_\mu^2 = n_0^2 - n_1^2 - \ldots - n_{D+1}^2 = -L^2 < 0 \,.
 }
We will present some evidence that the \NLsM\ \SAdS, as well as supersymmetric generalizations of it, has a fixed point in $d=2$
when $\alpha' D / L^2$ is close to $1$, at least for sufficiently large $D$.  Briefly, the evidence is this: the most accurate
calculations that we know, both as an expansion to fixed order in
$\alpha'/L^2$, and as an expansion in $1/D$ with finite $\alpha' D /
L^2$, lead to beta functions which have non-trivial zeroes.  These
zeroes arise because of competition between the one-loop term and
higher loop terms.  In the language of an effective action on target
space, the zeroes arise because of competition between the
Einstein-Hilbert term and higher powers of the curvature.

It is possible that this evidence is misleading.  Higher order
contributions to the beta function, both in an $\alpha'/L^2$
expansion and in a $1/D$ expansion, could be as large or larger than
the ones that we are able to compute.

There are various reasons to be interested in the behavior of the
beta function for the model \SAdS\ and its supersymmetric relatives.
Enormous interest has attached to backgrounds of string theory
involving $AdS_{D+1}$ factors because of their relation to conformal
field theories in $D$ dimensions \cite{juanAdS,gkPol,witHolOne} (for
a review see \cite{MAGOO}).  \NLsM's with non-compact symmetries
also arise in the treatment of disordered systems: see for example
\cite{WegnerMobility,Efetov:1997fw}.  And as we will see,
$AdS_{D+1}$ \NLsM's have some intrinsic interest, because although
they are related via a simple analytic continuation to the $O(N)$
model, they exhibit more complicated behavior that challenges some
of our usual field theory intuitions.

The bulk of this paper is devoted to an exposition of
two methods of computing the beta function for the theory \SAdS\ and
its supersymmetrizations.  In section~\ref{FIXEDORDER} we review
results at fixed order in $\alpha'$, {\it i.e.}~fixed loop order.
The state of the art is four loops.  In section~\ref{LARGED} we
explain how the leading $D$ dependence (and, in the bosonic case, the first sub-leading $D$ dependence) of all higher loop terms can be extracted from a $1/D$ expansion with finite $\alpha' D/L^2$.  There the state of the art is terms of order
$1/D^2$ relative to the one loop term.  We note a peculiar
feature of the beta function: its slope is large and negative at its
non-trivial zero, so corrections to scaling are controlled by an
operator of {\it negative} dimension.  We offer a heuristic
explanation of what this could mean, hinging on the supposition that
infrared fluctuations are large.  We also estimate the central charge of the non-trivial fixed point.  In section~\ref{DISCUSSION}, we
discuss possible consequences of a zero of the beta function: in particular, based on the results for the central charge, we speculate that the supersymmetric $AdS_5$ \NLsM\ may provide a string-scale holographic dual of Yang-Mills theory, and that finite volume quotients of $AdS_3$ may be used in compactifications to four dimensions.  We conclude with a review of our results and our conjectures in section~\ref{CONCLUSIONS}.

\section{Anti-de Sitter target spaces at fixed order in $\alpha'$}
\label{FIXEDORDER}

\begin{figure}
  \centerline{\includegraphics{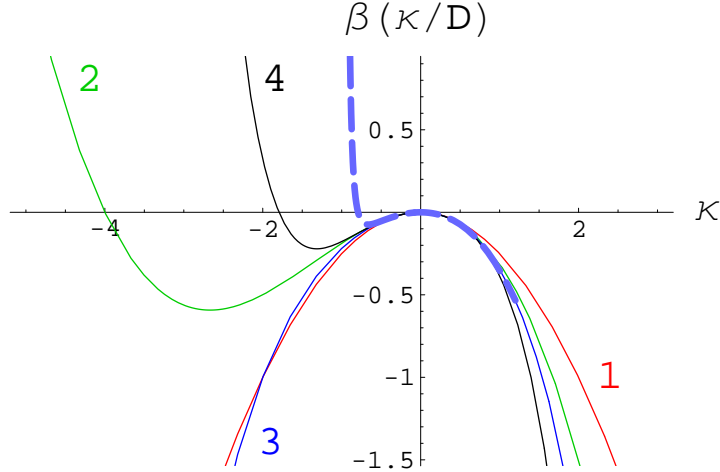}}
  \caption{The solid lines show successive fixed-order approximations to $\beta(g)$ for $D=4$.  For comparison with large $D$ results, it is convenient to use $\kappa = gD$ to parametrize the horizontal axis.  Each approximation is marked with its loop order.  The dashed line is the large $D$ result through order $D^{-2}$, also for $D=4$.}\label{FixedOrderBeta}
 \end{figure}

The partition function of the \NLsM\ \SAdS\ depends on $\alpha'$ and $L^2$ only in the combination
 \eqn{gDef}{
  g = -{\alpha' \over L^2} \,,
 }
which we define to be negative in order to anticipate a connection with the literature on $O(N)$ vector models.  In the scheme of dimensional regularization with minimal subtraction, one obtains the following beta function, up to four loops \cite{WegnerFourLoop, Jack1, Jack2}:
 \eqn{BosonicBeta}{
  \beta(g) &= -D g^2 - D g^3 - {1 \over 4} D(D+4) g^4  \cr
    &\qquad{} +
   \left( {D^3 \over 12} - {3 \over 2} (1+\zeta(3)) D^2 +
    {1 \over 2} (3\zeta(3)-1) D \right) g^5 + O(g^6) \,.
 }
As is evident from figure~\ref{FixedOrderBeta}, the non-trivial
fixed point is present or absent depending on how many terms one
retains.  This is discouraging.  At first glance, it seems not
merely plausible but likely that a computation of the $O(g^6)$ term
would make the non-trivial zero disappear.  However, as we shall
describe in section~\ref{LARGED}, merging the fixed order
information \BosonicBeta\ with the best results we could obtain from
a large $D$ expansion, one winds up with a beta function that does
have a non-trivial fixed point.  The large $D$ results contain
partial information about terms in $\beta(g)$ with arbitrarily high
powers of $g$.

Before presenting results from a large $D$ expansion, we will describe the relation of the $AdS_{D+1}$ \NLsM\ to the $O(N)$ vector model with $N=D+2$, present the supersymmetrizations of the $AdS_{D+1}$ \NLsM, and briefly survey results on the beta functions of each model.

\subsection{The $O(N)$ vector model continued to negative coupling}
\label{NEGATIVECOUPLING}

The partition function for the bosonic $AdS_{D+1}$ \NLsM\ can be written in a variety of ways:
 \eqn{AdSpartition}{
  Z &= \int {\cal D} n_\mu \, \delta(n_\mu^2 + L^2) \theta(n_0)
    \exp\left\{ -{1 \over 4\pi\alpha'} \int d^d x \,
     (\partial n_\mu)^2 \right\}
      \cr
    &= \int {{\cal D} \vec{n} \over \sqrt{L^2 + \vec{n}^2}}
     \exp\left\{ -{1 \over 4\pi\alpha'} \int d^d x \,
     \left[ (\partial\vec{n})^2 - (\partial\sqrt{L^2+\vec{n}^2})^2
      \right] \right\}  \cr
    &= \int {{\cal D} \vec\Pi \over \sqrt{1 - g \vec\Pi^2}}
    \exp\left\{-{1 \over 4\pi} \int d^d x \,
      \left[ (\partial\vec\Pi)^2 +
      {1 \over g} \left(
       \partial\sqrt{1-g \vec\Pi^2} \right)^2 \right]
       \right\}
 }
where we have split $n_\mu = (n_0,\vec{n})$ into the time-like component and the spatial $(D+1)$-component vector $\vec{n}$, defined $\vec\Pi = \vec{n}/\sqrt{\alpha'}$, dropped some inessential prefactors,\footnote{Dropping infinite prefactors is a formal manipulation, particularly since one of them is the reciprocal of the infinite volume of $AdS_{D+1}$.  To see why this factor should be there, consider defining the \NLsM\ on a finite lattice rather than in a continuum limit.  Then to get a finite partition function, one must fix one spin to a particular location in $AdS_{D+1}$.  See \cite{Duncan:2004ik} for a more thorough discussion.  This subtlety shouldn't affect the beta function.} and defined $g$ as in \gDef.  Recall that $g<0$.

Now recall the classic perturbative treatment of the low-temperature phase \cite{BrezinZJ}, in which one starts with an $N$-component Euclidean vector field $n_\mu$ subject to the constraint $n_\mu^2 = \delta^{\mu\nu} n_\mu n_\nu = L^2 > 0$ and writes the partition function as
 \eqn{SNpartition}{
  Z &= \int {\cal D} n_\mu \, \delta(n_\mu^2 - L^2)
   \exp\left\{ -{1 \over 4\pi\alpha'} \int d^d x \, (\partial n_\mu)^2
    \right\}  \cr
   &= \int {{\cal D} \vec{n} \over \sqrt{L^2 - \vec{n}^2}}
    \exp\left\{ -{1 \over 4\pi\alpha'} \int d^d x
     \left[ (\partial\vec{n})^2 + (\partial\sqrt{L^2 - \vec{n}^2})^2
      \right] \right\}  \cr
   &= \int {{\cal D} \vec\Pi \over \sqrt{1 - g \vec\Pi^2}}
    \exp\left\{-{1 \over 4\pi} \int d^d x \,
      \left[ (\partial\vec\Pi)^2 +
      {1 \over g} \left(
       \partial\sqrt{1-g \vec\Pi^2} \right)^2 \right]
       \right\}
 }
where we have again split $n_\mu = (n_0,\vec{n})$ into a single (Euclidean) component $n_0$ and a $(N-1)$-component vector $\vec{n}$.  $\vec\Pi$ is defined identically as in \AdSpartition, but now $g = \alpha'/L^2$.  Note that $g>0$.

The key observation is that the last lines of \AdSpartition\ and \SNpartition\ are identical.  The sign of $g$ determines whether one is treating the $S^{N-1}$ model or the $AdS_{D+1}$ model.\footnote{We will persist in using both $D$ and $N=D+2$ in order to ease the notational transition from well-known results on the $O(N)$ vector model to the ${\rm AdS}_{D+1}/{\rm CFT}_D$ correspondence.}  So, at the perturbative level, one may simply continue a quantity like $\beta(g)$ from the $O(N)$ model to negative $g$, set $N=D+2$, and apply the result to the $AdS_{D+1}$ \NLsM.  At a non-perturbative level, it is less clear that there is a definite relation between the $O(N)$ model and the $AdS_{D+1}$ model: the obvious difficulty in comparing the last lines of \AdSpartition\ and \SNpartition, for example, is that in the latter case one must explicitly bound $|\vec\Pi|^2 < 1/g$.  (More precisely, one must attach the lower hemisphere of $S^{N-1}$ to complete the partition function.)

\subsection{Supersymmetrizations of the $O(N)$ model}
\label{SUPERSYMMETRIZATIONS}

Having understood that results from the $O(N)$ model can be applied directly to the $AdS_{D+1}$ via the continuation discussed in the previous section, let us now re-express the $O(N)$ model and its supersymmetrizations in the form that is most convenient for perturbative calculations:
 \eqn{ZonceMore}{
  Z &= \int {\cal D} S \, \delta(S^2-1)
     \exp\left\{ -{1 \over 4\pi g} \int d^d x (\partial S)^2
       \right\}   \cr
    &= \int {\cal D} S {\cal D} \sigma
     \exp\left\{ -{1 \over 4\pi g} \int d^d x \Big[
      (\partial S)^2 + \sigma (S^2-1) \Big] \right\} \,,
 }
where $\sigma$ runs over imaginary values and we omit the target space
index on $S$.  The $(1,1)$ supersymmetric extension of the $O(N)$
model involves $N$ superfields ${\bf S}_\mu =
(S_\mu,\psi_\mu,F_\mu)$ and one additional superfield ${\bf \Phi} =
(\phi,u,\sigma)$ to enforce the constraint ${\bf S}^2 = 1$
\cite{gracey-short,gracey-long}:
 \eqn{OneOneComponents}{
  S = {1 \over 4\pi g} \int d^2 x \, \Bigg[
    (\partial S)^2 + \bar\psi i \slashed\partial \psi + F^2 +
    \sigma (S^2-1) + \phi \bar\psi \psi + 2 \bar{u} \psi S +
    2 S F \phi \Bigg] \,.
 }
The $(1,0)$ supersymmetric extension involves $N$ superfields ${\bf S_\mu} = (S_\mu,\psi_\mu)$ and an additional spinorial superfield ${\bf U} = (u,\sigma)$ to enforce the constraint:
 \eqn{OneZeroComponents}{
  S = {1 \over 4\pi g} \int d^2 x \, \Bigg[
    (\partial S)^2 + \bar\psi i \slashed\partial \psi +
    \sigma (S^2-1) + 2 \bar{u} \psi S \Bigg] \,.
 }
$\psi_\mu$ and $u$ are both chiral, but with opposite handedness.  In all cases, $g = \alpha'/L^2$ where $L$ is the radius of $S^{N-1}$.  Appropriate continuations of \OneOneComponents\ and \OneZeroComponents\ to negative $g$ lead to supersymmetric $AdS_{D+1}$ \NLsM's, analogously to the treatment in section~\ref{NEGATIVECOUPLING} of the bosonic case.

One can regard \ZonceMore, \OneOneComponents, and \OneZeroComponents\ as the starting points for describing bosonic, type~II, and heterotic strings propagating on a sphere $S^{N-1}$ with radius $L = \sqrt{\alpha'/g}$.  These are not consistent string backgrounds, but one may nevertheless borrow from the literature \cite{friedan, alvarez-gaume, gross-witten, grisaru-long, grisaru-short, callan} on general \NLsM's on the string worldsheet to extract fully covariant forms of the beta function.  In a minimal subtraction scheme, we have
\eqn{LeadingCorrection}{\seqalign{\span\TT\qquad & \span\TR}{
 bosonic: & \beta_{ij} = \alpha' R_{ij} + {\alpha'^2 \over 2}
   R_{iklm} R_j{}^{klm} + O(\alpha'^3)  \cr
 heterotic: & \beta_{ij} = \alpha' R_{ij} + {\alpha'^2 \over 4}
   R_{iklm} R_j{}^{klm} + O(\alpha'^3) \cr
 \hbox{type~II:} & \beta_{ij} = \alpha' R_{ij} + \frac{\zeta(3)
\alpha'^4}{2} R_{mhki}R_{jrt}{}^m (R^k{}_{qs}{}^r R^{tqsh} +
R^k{}_{qs}{}^t R^{hrsq}) + O(\alpha'^5) \,.  }} To obtain \LeadingCorrection, we have set to zero all deformations corresponding to matter fields (for example, $B_{ij}$), and assumed $R_{ijkl;m} = 0$, as is appropriate for any symmetric space.  The beta function for $g$ may be expressed as
 \eqn{ExpressBetaBeta}{
  \beta(g) \equiv M {\partial g \over \partial M} =
    -{g \over N-1} G^{ij} \beta_{ij} \,.
 }
Plugging
 \eqn{RforS}{
  R_{ijkl} = {1 \over L^2} (g_{ik} g_{jl} - g_{jk} g_{il})
 }
into \LeadingCorrection, and using \ExpressBetaBeta, one obtains
 \eqn{SeveralBetas}{\seqalign{\span\TT\qquad & \span\TR}{
 bosonic: & \beta(g) = -D g^2 - D g^3 + O(g^4) \cr
 heterotic: & \beta(g) = -D g^2 - {1 \over 2} D g^3 + O(g^4)  \cr
 \hbox{type~II:} & \beta(g) = -D g^2 -
   {3 \over 2} \zeta(3) D(D-1) g^5 + O(g^6) \,,
 }}
where we have expressed the final results in terms of $D = N-2$ to facilitate comparison with \BosonicBeta.  Evidently, the bosonic result agrees with \BosonicBeta,\footnote{Note that we do not even need to appeal to the continuation argument of section~\ref{NEGATIVECOUPLING} to relate the bosonic result in \LeadingCorrection\ to \BosonicBeta: we could plug the negative curvature metric of $AdS_{D+1}$ into the covariant expressions directly and wind up with the same two-loop result.} and all three beta functions to the order specified have non-trivial zeroes for negative $g$.

Let us now briefly anticipate the gist of section~\ref{LARGED}.  The
fixed-order results so far sketched for the bosonic \NLsM\ and its
$(1,1)$ supersymmetrization can be supplemented by the leading-order
$D$ dependence of the coefficients of all higher powers in $g$ from
a large $D$ expansion, obtained for the most part from
\cite{Vone,Vtwo} in the bosonic case and \cite{gracey-short,
gracey-long} in the $(1,1)$ supersymmetric case. The results will be
beta functions with zeroes for $gD$ slightly larger than $-1$. To
our knowledge, no large $D$ results exist in the literature for the
heterotic case; this is an avenue for future explorations.

\subsection{Scheme dependence}
\label{SCHEME}

It is well known (see for example \cite{gross-sloan}) that field redefinitions can alter terms in the beta function at two loops and higher so that they involve only the Weyl tensor.  Such redefinitions do not affect the S-matrix elements that were used \cite{gross-witten} to anticipate the existence of an $\alpha'^3$ term in the beta function for the general $(1,1)$ supersymmetric \NLsM.

$AdS_{D+1}$ has no Weyl curvature, so we can conclude for the
$AdS_{D+1}$ \NLsM\ (and its supersymmetrizations) that there is a
scheme---call it a ``Weyl tensor'' scheme---where higher loop terms
make no contribution to the beta function at all.  If a Weyl tensor
scheme is employed, then clearly there can be no non-trivial zero of
the beta function.\footnote{A preference for the scheme that leads
to dependence of higher loop terms only on the Weyl tensor stems
from the fact that there is an on-shell superspace formulation of
type~IIB supergravity \cite{Howe:1983sr} in which the Weyl tensor
rather than the Riemann tensor enters into the superfield for
linearized perturbations (in other words, the on-shell graviton
superfield).  Unless one can show that an off-shell formulation in
which supersymmetry requires using the Weyl tensor in place of the
Riemann tensor, we do not see any compelling reason to choose a Weyl
scheme even in the $(1,1)$ supersymmetric case.}

Existence or non-existence of a fixed point is supposed not to
depend on scheme.  But what could happen is that a Weyl scheme
pushes the zero to $g \to -\infty$.  In the absence of a systematic
treatment of the stability of various schemes, it seems best to us
to choose a standard one (like dimensional regularization with
minimal subtraction) that does not presuppose an answer to the
question we're interested in---as a Weyl scheme effectively does for
the existence or non-existence of a non-trivial $AdS_{D+1}$ fixed
point.

\section{Anti-de Sitter target spaces in a $1/D$ expansion}
\label{LARGED}

It is apparent from \BosonicBeta\ that the effective expansion parameter is $gD$ rather than $g$.  More precisely: except for the one-loop term, the coefficient of the $O(g^n)$ term in $\beta(g)$ is a polynomial in $D$ of order $n-2$.  In the $(1,1)$ supersymmetric case displayed in \SeveralBetas, the coefficient of the $O(g^5)$ term is only quadratic in $D$---one power less than the corresponding coefficient in \BosonicBeta.

These observations can be systematized through sophisticated large $D$ techniques pioneered by Vasiliev et al \cite{Vone, Vtwo} for the bosonic $O(N)$ \NLsM\ and extended by Gracey \cite{gracey-short, gracey-long} to the $(1,1)$ supersymmetric case.  For our purposes, the first significant claim (see \cite{gracey-short, gracey-long}) is that in
 \eqn{HowManyDims}{
  d=2\mu=2+\epsilon
 }
worldsheet dimensions, one may express
 \eqn{betaExpress}{
  {\beta(g) \over g} = \epsilon - \kappa +
   {1 \over D} b_1(\kappa) + {1 \over D^2} b_2(\kappa) +
    O(D^{-3}) \,,
 }
where $\kappa = gD$ and $O(D^{-3})$ means $1/D^3$ times a function of $\kappa$ only.  The functions $b_n(\kappa)$ have a power series expansion around $\kappa=0$ whose first term is at least order $\kappa^{n+1}$.

The second significant claim is that, in a minimal subtraction scheme, there is no $\epsilon$ dependence except as shown explicitly in \betaExpress.  To see this, recall that in such a scheme, the bare coupling $g_B$ and the renormalized coupling $g_R$ are related by
 \eqn{BareVersusR}{
  g_B = M^{-\epsilon} \sum_{k=0}^\infty
   {a_k(g_R) \over \epsilon^k} \,,
 }
where $a_0(g_R) = g_R$ and the remaining $a_k(g_R)$ are power series in $g_R$ whose coefficients do not depend on $M$ or $\epsilon$.  The left side of \BareVersusR\ is obviously independent of $M$.  In order for the right hand side to be independent of $M$, one must have
 \eqn{BetaImpose}{
  M {\partial g_R \over \partial M} \equiv \beta(g_R)
    = \epsilon g_R + \left( 1 - g_R {\partial \over \partial g_R}
     \right) a_1(g_R) \,.
 }
Evidently, $\beta(g)/g$ depends on $\epsilon$ only additively, as
indicated in \betaExpress.\footnote{Renormalizability requires that
the higher $a_k$ satisfy so-called pole equations, which accounts
for the fact that $\beta(g_R)$ depends only on $a_1$:
 \eqn{PoleEqs}{
  \left( 1 - g_R {\partial \over \partial g_R} \right)
   a_{k+1}(g_R) &= \left( 1 - g_R {\partial \over \partial g_R}
     \right) a_1(g_R) {\partial \over \partial g_R} a_k(g_R) \,.
 }}

For $\epsilon$ greater than $0$ but less than some finite upper bound, there is a non-trivial fixed point at a positive $\kappa_c$ satisfying
 \eqn{KappaExpress}{
  \epsilon = \kappa_c - {1 \over D} b_1(\kappa_c) -
    {1 \over D^2} b_2(\kappa_c) + O(D^{-3})
 }
The leading corrections to power law scaling near the fixed point are determined by the slope of the beta function, i.e.~in terms of the quantity
 \eqn{betagc}{
  \lambda = -{1 \over 2} \beta'(g_c)
 }
where of course $g_c = \kappa_c / D$, and the derivative of $\beta(g)$ is taken with $\epsilon$ held fixed.  The third main claim of \cite{Vone,Vtwo} is that position space diagrammatic techniques allow an independent determination of $\lambda$ as a power series in $1/D$:
 \eqn{lambdaExpress}{
  \lambda &= \lambda_0(\epsilon) +
     {\lambda_1(\epsilon) \over D} + {\lambda_2(\epsilon) \over D^2}
     + O(D^{-3}) \,,
 }
where the functions $\lambda_i(\epsilon)$ in \lambdaExpress\ have
been computed explicitly.

Given \betaExpress-\lambdaExpress, one can determine the $b_i$ in terms of the $\lambda_i$.  Here is the algebra: first one uses \KappaExpress\ to rewrite \lambdaExpress\ as
 \eqn{lEagain}{
  \lambda &= \lambda_0(\kappa_c) + {1 \over D} \Big[
      \lambda_1(\kappa_c) - b_1(\kappa_c) \lambda_0'(\kappa_c)
       \Big]  \cr
   &\qquad{}+ {1 \over D^2} \left[
      \lambda_2(\kappa_c) - b_2(\kappa_c) \lambda_0'(\kappa_c) -
       b_1(\kappa_c) \lambda_1'(\kappa_c) + {1 \over 2}
        b_1(\kappa_c)^2 \lambda_0''(\kappa_c) \right] + O(D^{-3}) \,.
 }
Next one computes $\beta'(g_c)$ in terms of the series expansion \betaExpress:
 \eqn{bEagain}{
  -{1 \over 2} \beta'(g_c) &= -{\epsilon \over 2} + \kappa_c -
    {1 \over 2D} \Big[ b_1(\kappa_c) + \kappa_c b_1'(\kappa_c) \Big] -
    {1 \over 2D^2} \Big[ b_2(\kappa_c) + \kappa_c b_2'(\kappa_c) \Big]
     + O(D^{-3})  \cr
   &= {\kappa_c \over 2} - {\kappa_c b_1'(\kappa_c) \over 2D} -
     {\kappa_c b_2'(\kappa_c) \over 2D^2} + O(D^{-3}) \,,
 }
where in the second line we have again used \KappaExpress\ to eliminate $\epsilon$.  Finally, one compares terms in the second lines of \lEagain\ and \bEagain\ to obtain
 \eqn{BfromL}{
  \lambda_0(\kappa_c) &= {\kappa_c \over 2}  \cr
  \lambda_1(\kappa_c) - b_1(\kappa_c) \lambda_0'(\kappa_c) &=
    -{1 \over 2} \kappa_c b_1'(\kappa_c)  \cr
  \lambda_2(\kappa_c) - b_2(\kappa_c) \lambda_0'(\kappa_c) -
    b_1(\kappa_c) \lambda_1'(\kappa_c) +
    {1 \over 2} b_1(\kappa_c)^2 \lambda_0''(\kappa_c) &=
    -{1 \over 2} \kappa_c b_2'(\kappa_c) \,.
 }
The equations \BfromL\ hold for any $\kappa_c$ greater than $0$ and less than some finite upper bound: thus the second and third can be regarded as differential equations for $b_1$ and $b_2$.  They may be integrated to obtain
 \eqn{IntegratedBs}{
  b_1(\kappa) = -2\kappa \int_0^\kappa d\xi \,
    {\lambda_1(\xi) \over \xi^2} \qquad
  b_2(\kappa) = -2\kappa \int_0^\kappa d\xi \,
    {\lambda_2(\xi) - b_1(\xi) \lambda_1'(\xi) \over \xi^2} \,.
 }
It is not obvious from what we have summarized so far that the lower limits of the integrals in \IntegratedBs\ should be $0$.  This will become clear once the explicit expressions for the $\lambda_i$ are in hand.

It is to be emphasized that $b_1(\kappa)$ and $b_2(\kappa)$ are defined through \IntegratedBs\ for finite $\kappa$.  More precisely: the treatment \betagc-\IntegratedBs\ defines $b_1(\kappa)$ and $b_2(\kappa)$ directly on some interval starting at $0$ and extending to finite positive values of $\kappa$.  Through analytic continuation, as in section~\ref{NEGATIVECOUPLING}, we extract the beta function of the $AdS_{D+1}$ \NLsM\ through order $1/D^2$, again for finite $\kappa = -\alpha' D / L^2$.  As we shall describe in section~\ref{SINGULARITY}, no singularities are encountered in this analytic continuation until $\kappa=-3$ for $b_1(\kappa)$ and $\kappa=-1$ for $b_2(\kappa)$.  

The critical exponent $\lambda$ is (in principle) a measurable quantity pertaining to the non-trivial fixed point in $d=2+\epsilon$ dimensions of the $O(N)$ \NLsM, with $D=N-2$.  As such, it doesn't suffer from any scheme ambiguities.  There is clearly a certain attractiveness in the strategy of folding all the difficult diagrammatic calculations into a determination of $\lambda$ (as well as other critical exponents) and then extracting the beta function in a minimal subtraction scheme in the very last step.

Having summarized all but the difficult calculations, we will turn in section~\ref{BOSONIC} to the promised determinations of $\lambda_1(\epsilon)$ and $\lambda_2(\epsilon)$ in the bosonic \NLsM.  The final results can be previewed in figure~\ref{FixedOrderBeta}.  As promised, there is once again a non-trivial fixed point for negative $\kappa$, corresponding, apparently, to a CFT with target space $AdS_{D+1}$.  The reader who wishes to skip the technical details can find the main results in equations~\eno{AnomCoefs} and~\eno{GotDeltasSecond}, supplemented by the definitions~\eno{AnomExpand}, \eno{SeveralDefs},and~\eno{Ks}.

\subsection{Explicit results for the bosonic case}
\label{BOSONIC}

In this section, we will give a fairly complete summary of the
bosonic calculation through $O(D^{-2})$ \cite{Vone, Vtwo} for three
reasons: first, the position space techniques employed are less well
known than fixed order perturbative techniques; second, we find a
minor discrepancy in the final result of \cite{Vtwo}, at least as it
appears in translation as cited; and third, we will identify one
particular three loop diagram which is almost entirely responsible
for the effects we are interested in.

\begin{figure}
  \centerline{\includegraphics[width=3.5in]{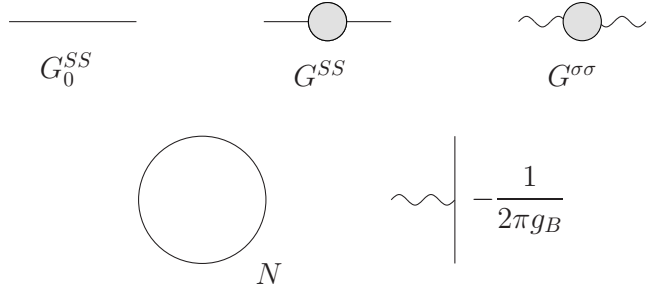}}
  \caption{Feynman rules for the $O(N)$ model \ZWGamma.  Shaded circles indicate a dressed propagator.  There is no undressed propagator for $\sigma$.  There is a tadpole for $\sigma$ which we omit because it does not contribute to the calculations of interest.  A loop of $S_\mu$ picks up a factor of $N$ whether the propagators in it are dressed or undressed, and regardless of how many $SS\sigma$ vertices it may include.}\label{FeynmanRules}
 \end{figure}

To fix notation, consider the following generating functionals:
 \def\extremum{\mathop{\rm extremum}}
 \eqn{ZWGamma}{
  Z[J_S,J_\sigma] &= \int {\cal D} S {\cal D} \sigma \,
   \exp\left\{ -{1 \over 4\pi g_B} \int d^d x \,
    \left[ (\partial S)^2 + \sigma (S^2 - 1) \right] +
    \int d^d x \left[ J_S S + J_\sigma \sigma \right] \right\}  \cr
   &= e^{W[J_S,J_\sigma]}
   = \extremum_\sigma \exp\left\{ -\Gamma[S,\sigma] +
    \int d^d x \left[ J_S S + J_\sigma \sigma \right] \right\}
 }
The connected Green's functions $G^{(n_S,n_\sigma)}$ are derivatives
of $W$, i.e.~the connected $(n_S+n_\sigma)$-point function $\langle
S(x_1) \cdots S(x_{n_S}) \sigma(y_1) \cdots \sigma(y_{n_\sigma})
\rangle_c$.  The 1PI Green's functions $\Gamma^{(n_S,n_\sigma)}$ are
analogous derivatives of $\Gamma$.  For two-point functions we will
adopt notations like $G^{SS}$ instead of $G^{(2,0)}$.  Near an
ultraviolet-stable fixed point, the scaling parts of $G^{SS}$ and
$G^{\sigma\sigma}$ and the leading correction to them may be
expressed for small but non-zero $x$ as
 \eqn{ScalingParts}{
  G^{SS}(x) = {C^{SS} \over x^{2\Delta_S}} \big[ 1 +
   \tilde{C}^{SS} x^{2\lambda} + \ldots \big] \qquad
  G^{\sigma\sigma}(x) = {C^{\sigma\sigma} \over x^{2\Delta_\sigma}} \big[ 1 +
   \tilde{C}^{\sigma\sigma} x^{2\lambda} + \ldots \big] \,,
 }
where as noted above, $2\lambda = -\beta'(g_c)$ at the UV critical
point.  We have omitted to write the tensor structure of $G^{SS}$:
it is proportional to $\delta_{\mu\nu}$.  Indeed, because both the
bare propagator for $S_\mu$ and the bare $SS\sigma$ vertex are
$\delta_{\mu\nu}$ times a scalar function, all Green's functions
$G^{(n_S,n_\sigma)}$ or $\Gamma^{(n_S,n_\sigma)}$ may be expressed
as scalar functions times symmetrized products of $\delta_{\mu_i
\mu_j}$ where $i,j = 1,2,\ldots,n_S$: in particular, no factors like
$(x_1-x_2)^{\mu_i}$ can ever appear.  Tensor structure may be
completely ignored for the calculations of interest to us; the only
rule to remember is that every loop of $S_\mu$ picks up a factor of
$N$.  See figure~\ref{FeynmanRules} for the Feynman rules that we
will use.

\begin{figure}
  \centerline{\includegraphics[width=6.5in]{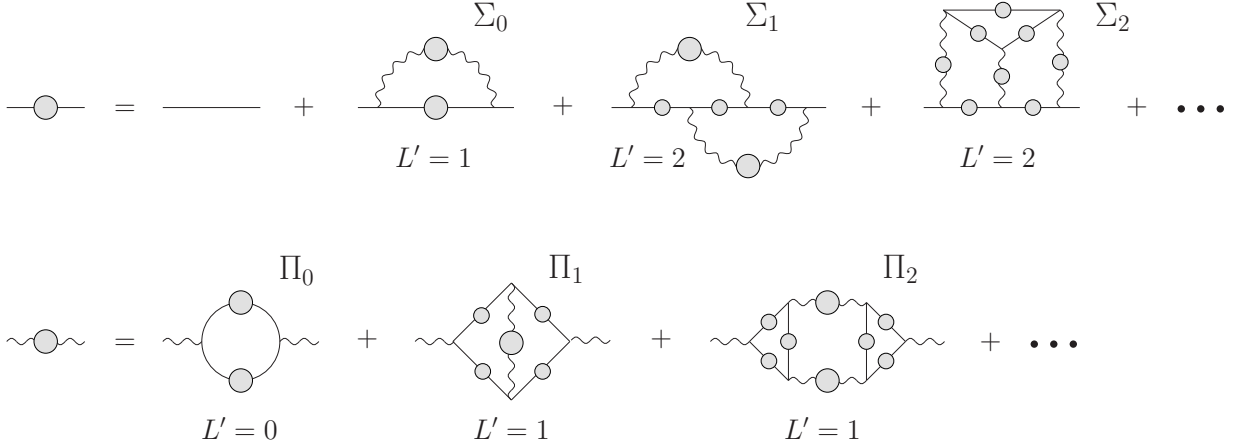}}
  \caption{Graphical representation of the Dyson equations for $G^{SS}$ and $G^{\sigma\sigma}$.  All graphs have external legs amputated, so (for instance) the left hand sides are $\Gamma^{SS}$ and $\Gamma^{\sigma\sigma}$.  The effective loop order $L'$, defined as the number of loops minus the number of loops of $S_\mu$, is indicated for each graph.  See the beginning of section~\ref{NEXTORDER} for a more complete discussion of effective loop order.  The graph labeled $\Pi_2$ makes the crucial contribution leading to a non-trivial zero of the beta function; see section~\ref{SINGULARITY}.}\label{DysonEqs}
 \end{figure}

The renormalized dimensions of the operators in this theory can be expanded in $1/D$:
 \eqn{AnomExpand}{
  \Delta_S &= \Delta_{S0} + {\Delta_{S1} \over D} +
    {\Delta_{S2} \over D^2} + \ldots  \cr
  \Delta_\sigma &= \Delta_{\sigma0} + {\Delta_{\sigma1} \over D} +
    {\Delta_{\sigma2} \over D^2} + \ldots  \cr
  \lambda &= \lambda_0 + {\lambda_1 \over D} + {\lambda_2 \over D^2}
    + \ldots \,.
 }
The method of \cite{Vone,Vtwo} is to self-consistently determine the
$\Delta_S$, $\Delta_\sigma$, and $\lambda$ to some order in $1/D$ by
plugging \ScalingParts\ and \AnomExpand\ into Dyson equations which
are represented graphically in figure~\ref{DysonEqs}.  The graphs
contributing to the Dyson equations are precisely the 1PI graphs,
with the exception of graphs containing subgraphs that already
appear at a lower efffective loop order.  So, for example, the
two-loop rainbow graph correction to $\Gamma^{SS}$ is omitted from
the Dyson equations because it would be generated by iterating the
Dyson equation for $G^{SS}$ truncated after the term labeled
$\Sigma_0$.  The undressed amputated $SS$ propagator is a
distribution supported at $x=0$, so it makes no contribution to the
scaling form \ScalingParts.  Thus the first term in the Dyson
equation for $G^{SS}$ could have been omitted.  Graphs with $\sigma$
tadpoles also do not contribute to the scaling forms \ScalingParts,
so we have entirely suppressed them.

\subsubsection{Lowest order results}
\label{LOWESTORDER}

The lowest order treatment is to include only the graphs $\Sigma_0$ and $\Pi_0$ in the Dyson equations:
 \eqn{LeadingEqns}{
  \Gamma^{SS}(x) &= \Sigma_0 = -\left( {1 \over 2\pi g_B} \right)^2
    G^{SS}(x) G^{\sigma\sigma}(x)  \cr
  \Gamma^{\sigma\sigma}(x) &= \Pi_0 = -{N \over 2}
   \left( {1 \over 2\pi g_B} \right)^2 G^{SS}(x)^2 \,.
 }
(The overall minus signs on the right hand sides are due to the sign
on $\Gamma[S,\sigma]$ in \ZWGamma.)  Explicitly,
 \eqn{xSpaceConstraints}{
  {p(\Delta_S) \over C^{SS}} {1 \over x^{2(2\mu-\Delta_S)}}
    \big[
     1 - \tilde{C}^{SS} q(\Delta_S,\lambda) x^{2\lambda}
    \big] &=
   -\left( {1 \over 2\pi g_B} \right)^2 {C^{SS} C^{\sigma\sigma} \over
     x^{2(\Delta_S + \Delta_\sigma)}} \big[ 1 + (\tilde{C}^{SS} +
      \tilde{C}^{\sigma\sigma}) x^{2\lambda} \big]  \cr
  {p(\Delta_\sigma) \over C^{\sigma\sigma}}
   {1 \over x^{2(2\mu-\Delta_\sigma)}} \big[ 1 - \tilde{C}^{\sigma\sigma}
     q(\Delta_\sigma,\lambda) x^{2\lambda} \big] &=
   -{N \over 2} \left( {1 \over 2\pi g_B} \right)^2
    {(C^{SS})^2 \over x^{4\Delta_S}} \big[ 1 + 2\tilde{C}^{SS}
      x^{2\lambda} \big]
 }
where $\mu = d/2 = 1+\epsilon/2$, as in \HowManyDims, and
 \eqn{SeveralDefs}{
  a(\Delta) \equiv {\Gamma(\mu-\Delta) \over \Gamma(\Delta)} \qquad
  p(\Delta) \equiv {a(\Delta-\mu) \over \pi^{2\mu} a(\Delta)} \qquad
  q(\Delta,\lambda) = {a(\Delta-\lambda) a(\Delta+\lambda-\mu) \over
    a(\Delta) a(\Delta-\mu)} \,.
 }
The factors of $p$ and $q$ in \xSpaceConstraints\ come from the inverse propagators $\Gamma$, which one finds by passing to momentum space using a standard Fourier integral:
 \eqn{StandardFourier}{
  \int d^d x {e^{-i k \cdot x} \over x^{2\Delta}} =
    {\pi^\mu a(\Delta) 2^{2(\mu-\Delta)} \over k^{2(\mu-\Delta)}}
     \,,
 }
and noting that in Fourier space, the inverse propagator is simply the algebraic inverse.

The equations \xSpaceConstraints\ imply
 \eqn{ParseXPC}{
  2\Delta_S + \Delta_\sigma = 2\mu \qquad
  p(\Delta_S) = -z = {2 \over N} p(\Delta_\sigma) \qquad
  z \equiv {(C^{SS})^2 C^{\sigma\sigma} \over (2\pi g_B)^2}
 }
from matching the leading powers of $x$, and
 \eqn{ParseTildes}{
  \begin{pmatrix} 1 + q(\Delta_S,\lambda) & 1 \\
   2 & q(\Delta_\sigma,\lambda) \end{pmatrix}
  \begin{pmatrix} \tilde{C}^{SS} \\ \tilde{C}^{\sigma\sigma}
   \end{pmatrix} = 0
 }
from matching the subleading powers.  In order for \ParseTildes\ to
admit a solution with non-zero $\tilde{C}^{SS}$ and
$\tilde{C}^{\sigma\sigma}$, the determinant of the matrix must
vanish, which is to say
 \eqn{Solvability}{
  \big[ q(\Delta_S,\lambda) + 1 \big] q(\Delta_\sigma,\lambda) = 2 \,.
 }
Once $\mu$ and $N=D+2$ are specified, \ParseXPC\ and \Solvability\
can be solved straightforwardly for $\Delta_S$, $\Delta_\sigma$, and
$\lambda$ in the $1/D$ expansions \AnomExpand:
 \eqn{AnomCoefs}{\seqalign{\span\TL & \span\TR\qquad & \span\TL & \span\TR}{
  \Delta_{S0} &= \mu-1 &
   \Delta_{S1} &= -2 {a(2-\mu) a(\mu-1) \over a(2) \Gamma(\mu+1)}
     \cr
  \Delta_{\sigma0} &= 2 & &  \cr
  \lambda_0 &= \mu-1 &
   \lambda_1 &= -2\Delta_{S1} {(2\mu-1)(\mu-1) \over \mu-2}
 }}
and also
 \eqn{zExpansion}{
  z = z_0 + {z_1 \over D} + {z_2 \over D^2} + O(D^{-3})
 }
where
 \eqn{zCoefs}{
  z_0 = 0 \qquad z_1 = \Delta_{S1} {\Gamma(\mu+1) \Gamma(\mu-1) \over
    \pi^{2\mu}} \,.
 }

Although it is possible to obtain expressions for $\Delta_{S2}$,
$\Delta_{\sigma1}$, $\lambda_2$, and $z_2$ from \ParseXPC\ and
\Solvability, these coefficients also receive contributions from the
graphs $\Sigma_1$, $\Sigma_2$, $\Pi_1$, and $\Pi_2$, as we shall
summarize in the next section.

\subsubsection{Effects at order $D^{-2}$}
\label{NEXTORDER}

To determine the order in $1/D$ at which a given graph begins to
contribute, replace each propagator $G^{SS}$ by the overall
coefficient $C^{SS}$, and likewise replace $G^{\sigma\sigma}$ by
$C^{\sigma\sigma}$.  The ``scaling amplitude'' of the graph is then
some monomial in $C^{SS}$, $C^{\sigma\sigma}$, $1/g_B$, and $N$.  It
is straightforward to show that the scaling amplitudes of all the
graphs contributing to a given Dyson equation are some fixed
monomial times powers of $z$ and $D$, and that if one replaces $N$
by $D$ and $z$ with $1/D$---consistent with the scaling
\zCoefs---then the resulting power of $D$ is $D^{1-L'}$, where the
``effective loop order'' $L'$ is the number of loops minus the
number of loops of $S_\mu$.  For example, $\Pi_0$ has scaling
amplitude ${(C^{SS})^2 N \over g_B^2}$, whereas $\Pi_2$ has scaling
amplitude
 \eqn{PiTwoScaling}{
  {(C^{SS})^6 (C^{\sigma\sigma})^2 \over g_B^6} N^2 \sim
   {(C^{SS})^2 N \over g_B^2} z^2 N \sim
   {(C^{SS})^2 N \over g_B^2} {1 \over D} \,.
 }
In the first step of \PiTwoScaling\ we have used the definition of
$z$ and discarded factors of $2$ and $\pi$; in the second step we
have used \zExpansion\ and discarded further $O(1)$ factors.
Evidently, $\Pi_2$ is suppressed relative to $\Pi_0$ by a single
power of $D$, as the effective loop order leads us to expect.  In
short, each graph starts to contribute at order $D^{1-L'}$.  See
figure~\ref{DysonEqs} for the effective loop order of each graph.
The number of graphs increases quickly as one proceeds to higher
effective loop orders.

To evaluate $\Pi_1$, $\Pi_2$, $\Sigma_1$, and $\Sigma_2$ in position
space, one must integrate over internal vertices.  Certain
identities to facilitate these computations were developed in
\cite{Vone,Vtwo}; see also \cite{Vasilev:2004yr} for a systematic
exposition.  Infinities arise in these position space integrals if
one uses $2\Delta_S + \Delta_\sigma = 2\mu$, as obtained in
\ParseXPC. These infinities are naturally regulated: each of the
four amplitudes of interest can be expanded as singular term,
proportional to $1/(2\mu-2\Delta_S-\Delta_\sigma)$, plus terms that
remains finite as we take $2\Delta_S + \Delta_\sigma \to 2\mu$.  The
singular terms vanish provided we set\footnote{It is actually no
coincidence that $\Delta_{\sigma 1} = -2\lambda_1$---it is a
consequence of the fact that $\Delta_\sigma = d - 2\lambda$
\cite{Ma}.}
 \eqn{AlterDeltaPhi}{
  \Delta_{\sigma1} = 4\Delta_{S1} {(2\mu-1)(\mu-1) \over \mu-2} \,,
 }
and from the finite terms one eventually finds corrections to \ParseXPC, \ParseTildes, and \Solvability\ \cite{Vone,Vtwo}.  The corrections to \ParseXPC\ are
 \eqn{AwfulDeltas}{
  p(\Delta_S) + z + z^2 \Sigma_1' + z^3 N \Sigma_2' = 0 \qquad
  {2 \over N} p(\Delta_\sigma) + z + z^2 \Pi_1' + z^3 N \Pi_2' = 0
 }
where
 \eqn{SigmaPis}{
  \Sigma_1' &= \half \Pi_1' =  \frac{\pi^{2\mu} a(\Delta_S)^2a(\dbeta)}{\Gamma(\mu)} \left(B(\Delta_\sigma) -
    B (\Delta_S)\right)  \cr
  \Sigma_2' &= \frac{2\pi^{4\mu} a(\dalpha)^3 a(\dbeta)^3
   a(\mu + \dalpha - \dbeta)}{\Gamma(\mu)} \left(B(\Delta_\sigma) - B(\Delta_S)\right)  \cr
  \Pi_2' &= \frac{\pi^{4\mu} a(\dalpha)^3 a(\dbeta)^3
   a(\mu + \dalpha - \dbeta)}{\Gamma(\mu)} \left(4 B(\Delta_\sigma) - 3 B(\Delta_S) -
    B(\mu + \Delta_S - \Delta_\sigma)\right)
 }
and
 \eqn{Ks}{
  B(x) &= \psi(x) + \psi(\mu-x) \,.
 }
The quantities $\Sigma_1'$, $\Sigma_2'$, $\Pi_1'$, and $\Pi_2'$,
each a function only of $\mu$, are the coefficients of the leading
power of $x$ in the finite parts of the corresponding graphs, with
powers of $C^{SS}$, $C^{\sigma\sigma}$, $g_B$, and $N$ removed, as
the dependence on these parameters has already been extracted into
\AwfulDeltas.  We still define $z = {(C^{SS})^2 C^{\sigma\sigma}
\over (2\pi g_B)^2}$, as in \ParseXPC.

The corrections to \ParseTildes\ are
 \eqn{AwfulMatrix}{
  & \begin{pmatrix} T^{SS} & T^{S\sigma} \\ T^{\sigma S} & T^{\sigma\sigma}
   \end{pmatrix}
  \begin{pmatrix} \tilde{C}^{SS} \\ \tilde{C}^{\sigma\sigma}
   \end{pmatrix} = 0 \cr
  & T^{SS} = -p(\Delta_S) q(\Delta_S,\lambda) + z + z^2 \Sigma_{1S}' +
     z^3 N \Sigma_{2S}'  \cr
  & T^{S\sigma} = z + z^2 \Sigma_{1\sigma}' + z^3 N \Sigma_{2\sigma}'  \cr
  & T^{\sigma S} = 2z + z^2 \Pi_{1S}' + z^3 N \Pi_{2S}'  \cr
  & T^{\sigma\sigma} =
   -{2 \over N} p(\Delta_\sigma) q(\Delta_\sigma,\lambda) +
    z^2 \Pi_{1\sigma}' + z^3 N \Pi_{2\sigma}' \,,
 }
where
 \eqn{SigmaPiPrimes}{\seqalign{\span\TL & \span\TR\qquad &
   \span\TL & \span\TR}{
  \Sigma_{1S}' &= 2 D_1 + D_2 &
   \Sigma_{1\sigma} &= 2 D_3  \cr
  \Sigma_{2S}' &= 2 D_6 + 2 D_7 + D_8 &
   \Sigma_{2\sigma}' &= 2 D_9 + D_{10}  \cr
  \Pi_{1S}' &= 4 D_4 &
   \Pi_{2\sigma}' &=  D_5  \cr
  \Pi_{2S}' &= 4 D_{11} + 2 D_{12} &
   \Pi_{2\sigma}' &= 2 D_{13}
 }}
and, following the notation of \cite{Vone, Vtwo},
 \eqn{Dvalues}{
  D_1 &= - {\pi^{2\mu} \over (2-\mu)\Gamma(\mu)^2}  \cr
  D_2 &= {\pi^{2\mu} \over (2-\mu)^2\Gamma(\mu-1)^2}  \cr
  D_3 &= D_4 =
   {\pi^{2\mu}(\mu^2 - 3\mu + 1) \over (2-\mu)^2\Gamma(\mu)^2}  \cr
  D_5 &= {3\pi^{2\mu} \over (2-\mu)(2\mu-3)\Gamma(\mu-1)^2} R_1  \cr
  D_6 &= -{\pi^{4\mu}(\mu^2 - 3\mu + 1)\Gamma(1-\mu) \over
    (2-\mu)^3\Gamma(\mu)\Gamma(2\mu-3)}  \cr
  D_7 &= {\pi^{4\mu}\Gamma(2-\mu) \over
    (2-\mu)\Gamma(\mu-1)\Gamma(2\mu-2)}
   \left[\frac{2\mu-3}{(2-\mu)^2} + 3 R_1\right] \cr
  D_8 &= -{\pi^{4\mu} \Gamma(4-\mu) \over
    (2-\mu)^5 \Gamma(\mu-1)\Gamma(2\mu-4)}  \cr
  D_9 &= D_{11} = {\pi^{4\mu} \Gamma(1-\mu)(-2\mu^2 + 7\mu - 4) \over
    (2-\mu)^3 \Gamma(\mu)\Gamma(2\mu-3)}  \cr
  D_{10} &= D_{12} = {3\pi^{4\mu}\Gamma(3-\mu) \over
    (2-\mu)^3\Gamma(\mu-1)\Gamma(2\mu-2)} R_1  \cr
  D_{13} &= {\pi^{4\mu}\Gamma(2-\mu) \over
    2(2-\mu)^3\Gamma(\mu-1)\Gamma(2\mu-2)}
   \left[6R_1 - R_2 - R_3^2\right] \,,
 }
with\footnote{Note that \cite{Vone} includes two inconsistent
definitions for $R_3$---the one in the main text of the paper is
correct, whereas the definition in the appendix contains an error.}
 \eqn{Rs}{
  R_1 &= \psi'(\mu-1) - \psi'(1)  \cr
  R_2 &= \psi'(2\mu-3) - \psi'(2-\mu) - \psi'(\mu-1) +
    \psi'(1)  \cr
  R_3 &= \psi(2\mu-3) + \psi(2-\mu) - \psi(\mu-1) - \psi(1) \,.
 }
The equations \AwfulMatrix\ arise from comparing a subleading term
in $\Gamma^{SS}$ or $\Gamma^{\sigma\sigma}$ to its form obtained
from the right hand side of a Dyson equation.  To match powers of
$x$, each graph on the right hand side needs to have all but one of
its propagators set equal to their leading power law
behaviors---that is, $C^{SS}/x^{2\Delta_S}$ or
$C^{\sigma\sigma}/x^{2\Delta_\sigma}$---while the last propagator is
set equal to $C^{SS} \tilde{C}^{SS}/x^{2(\Delta_S-\lambda)}$ or
$C^{\sigma\sigma}
\tilde{C}^{\sigma\sigma}/x^{2(\Delta_\sigma-\lambda)}$.
$\Sigma_{1S}'$, a function only of $\mu$, denotes the coefficient of
the finite part of $\Sigma_1$, with one $S$ propagator replaced by
its subleading behavior, and with factors of $C^{SS}$,
$C^{\sigma\sigma}$, $g_B$, $N$, and $\tilde{C}^{SS}$ removed.  The
other expressions in \SigmaPiPrimes\ have analogous meanings.  Each
contribution $D_i$ arises from a particular choice of which
propagator to assign subleading behavior to.

The corrections to \Solvability, of course, are
 \eqn{DetEquation}{
  T^{SS} T^{\sigma\sigma} - T^{S\sigma} T^{\sigma S} = 0 \,.
 }

It is now straightforward though tedious to plug the expansions \AnomExpand\ and \zExpansion\ into \AwfulDeltas\ and \DetEquation\ and obtain coefficients of arbitrarily high orders in $1/D$.  The ones which do not suffer further corrections from higher order graphs are
 \eqn{GotDeltasSecond}{
  z_2 &= -2z_1 + \frac{z_1 \eta_1}{\mu-2}
      \Bigg( 4 + \frac{8}{\mu-2} + 14\mu - 6\mu^2   \cr
    &\qquad{} + 2\big[ B(\mu-2) - B(2-\mu) \big] +
      \mu(2\mu-3)\big[ B(3-\mu) + B(\mu-2) - 2B(2-\mu) \big]
        \Bigg)  \cr
  \Delta_{S2} &= -2\Delta_{S1} + 2\Delta_{S1}^2
     \Bigg( {8\mu^5 - 42\mu^4 + 65\mu^3 - 34\mu^2 + 8\mu - 4 \over
       2\mu(\mu-1)(\mu-2)^2} +
       {z_2 \over 2\Delta_{S1} z_1} \Bigg)  \cr
  \lambda_2 &= -2\lambda_1 + 2\Delta_{S1}^2
     {\mu(\mu-1) \over (2-\mu)^2} \Bigg(
      {2(-4\mu^4 + 12\mu^3 - 5 \mu^2 - 6\mu + 2) \over \mu(\mu - 1)}
      \big[ B(2-\mu) - B(\mu-1) \big]  \cr
    &\qquad{} + {2\mu(2\mu - 3)^2 \over 2-\mu}
      \big[ 6R_1 - R_2 - R_3^2 \big] +
      3\mu(8\mu - 11)R_1  \cr
    &\qquad{} + 12\mu^2 - 18\mu + 20 + \frac{6}{\mu} -
      \frac{2}{\mu^2} - \frac{10}{2 - \mu} +
      \frac{10}{\mu - 1} - \frac{3}{(\mu - 1)^2} \Bigg) \,.
 }
In comparing with \cite{Vtwo}, one needs to know that $\eta = 2(\Delta_S - \Delta_{S0})$, and that the series expansions employed there are in $1/N$ rather than $1/D$: for instance, $z = z_0 + z_1/N + \tilde{z}_2/N^2$, where $\tilde{z}_2 = z_2 + 2 z_1$.  Thus, to obtain the coefficient $\tilde{z}_2$ of $1/N^2$, one simply removes the first term from the right hand side of the first equation in \GotDeltasSecond.

\subsubsection{Consistency checks}
\label{CONSISTENCY}

The discrepancies we find with \cite{Vtwo} are some minor differences in $\Delta_{S2}$ and $\lambda_2$ and the analogous quantities quoted there.  A highly non-trivial check made in \cite{Vtwo} is to compare $\nu = 1/2\lambda$ with results obtained in $d=3$ (that is, $\mu=3/2$) by studying the high-temperature phase of the $O(N)$ vector model \cite{okabe}:
 \eqn{CheckOkabe}{
  \nu = \sum_{n=0}^\infty {\tilde\nu_n \over N^n} \qquad
   \tilde\nu_0 = 1 \,, \quad
   \tilde\nu_1 = -{32 \over 3\pi^2} \,, \quad
   \tilde\nu_2 = {32 \over \pi^4} 
     \left( {112 \over 27} - \pi^2 \right) \,.
 }
These expressions agree with the results one gets using \AnomCoefs\ and \GotDeltasSecond.  In particular,
 \eqn{nuTwoTilde}{
  \tilde\nu_2 = {\lambda_1^2 -\lambda_0 (\lambda_2 + 2\lambda_1) 
    \over 2\lambda_0^3} \,.
 }
The final expression in \cite{Vtwo} for $\nu_2$ leads to a result that differs from the one quoted in \CheckOkabe\ by a factor $(176+27\pi^2)/(-112+27\pi^2)$.  Thus we believe that our expressions are correct, and that the inaccuracies appearing in \cite{Vtwo} are typographical.\footnote{We have tried without success to contact the authors of \cite{Vone,Vtwo} in order to discover whether the errors might be in translation.}

With $\lambda_1$ and $\lambda_2$ in hand, we may return to \IntegratedBs\ to extract the beta function.  Series expansions
 \eqn{LambdaSeries}{
  \lambda_0 &= {\epsilon \over 2}  \cr
  \lambda_1 &= {\epsilon^2 \over 2} + {\epsilon^3 \over 4} -
    {\epsilon^4 \over 8} + O(\epsilon^5)  \cr
  \lambda_2 &= {5+9\zeta(3) \over 4} \epsilon^4 + O(\epsilon^5)
 }
lead immediately to
 \eqn{AgreesWithWegner}{
  \beta(g) &= \epsilon g -D g^2 - D g^3 - {1 \over 4} D(D+4) g^4  \cr
    &\qquad{} +
   \left( {D^3 \over 12} - {3 \over 2} (1+\zeta(3)) D^2
     \right) g^5 + O(g^6) \,,
 }
which agrees with \BosonicBeta\ except for a term scaling as $D g^5$.  This term corresponds to an $O(D^{-3})$ contribution to $\lambda$, so it is evidently excluded from the calculations in this section.

There appeared to be some arbitrariness in choosing the lower limits of integration in \IntegratedBs\ to be $0$.  This arbitrariness is removed by requiring agreement with the one- and two-loop terms in \BosonicBeta.\footnote{There is a small caveat: there are two other choices for the lower limit of integration on the integral that determines $b_1(\kappa)$.  Both are irrational, and choosing one of them instead of $0$ does not alter the higher loop terms.  For the $b_2(\kappa)$ integral, there appears to be one choice other than $0$, but it is less than $-1$, hence on the other side of the singularity of primary interest to us.}  Thus the non-trivial check in comparing \AgreesWithWegner\ to \BosonicBeta\ is the three- and four-loop terms.

\subsubsection{A singularity at $\epsilon=-1$}
\label{SINGULARITY}

Because $\Gamma(z)$ has all its singularities on the real axis, the
same is true of $\lambda_1(\epsilon)$.  The singularity of
$\lambda_1(\epsilon)$ closest to $\epsilon=0$ is at $\epsilon=-3$,
whereas the singularity of $\lambda_1(\epsilon)$ closest to
$\epsilon=0$ is at $\epsilon=-1$.  It's clear from \IntegratedBs\
that $b_1(\kappa)$ and $b_2(\kappa)$ have the same radii of
convergence around $\epsilon=0$ as $\lambda_1(\epsilon)$ and
$\lambda_2(\epsilon)$, respectively.  Near their respective
singularities, one can show that
 \eqn{SeveralSeries}{\seqalign{\span\TL & \span\TR \qquad &
    \span\TL & \span\TR}{
  \lambda_1(\epsilon) &= -{3/2\pi^2 \over \epsilon+3} +
    O(1) &
  b_1(\kappa) &= -{\log(3+\kappa) \over \pi^2} + O(1)  \cr
  \lambda_2(\epsilon) &= {32/3\pi^4 \over (\epsilon+1)^4} +
    O\big[ (\epsilon+1)^{-3} \big] &
  b_2(\kappa) &= -{64/9\pi^4 \over (\kappa+1)^3} +
    O\big[ (\kappa+1)^{-2} \big] \,.
 }}
All terms in $\lambda_2$ are regular at $\epsilon = -1$ (\ie, $\mu =
1/2$), except for $R_3$, which originates only from the diagram
$\Pi_2$ of figure \ref{DysonEqs}, upon replacing a $\sigma$
propagator with its leading correction to scaling (see the quantity
$D_{13}$ in \Dvalues).  Evidently, this diagram is somehow
responsible for the existence of the fixed point at this order.

\begin{figure}
  \centerline{\includegraphics[width=3.2in]{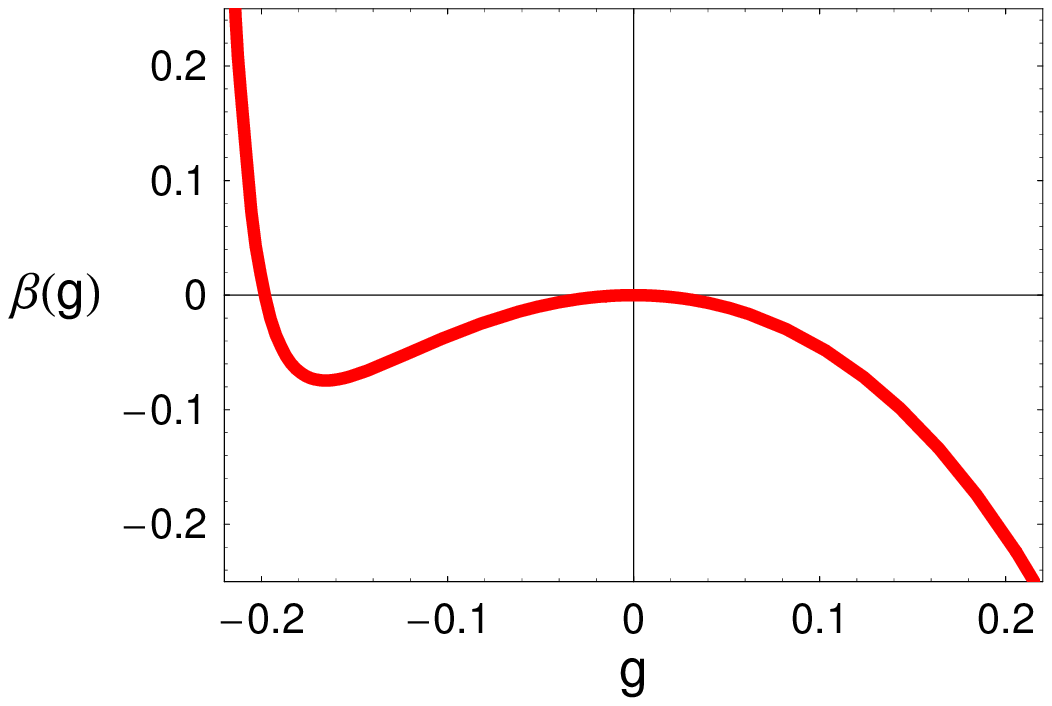}
              \includegraphics[width=3.2in]{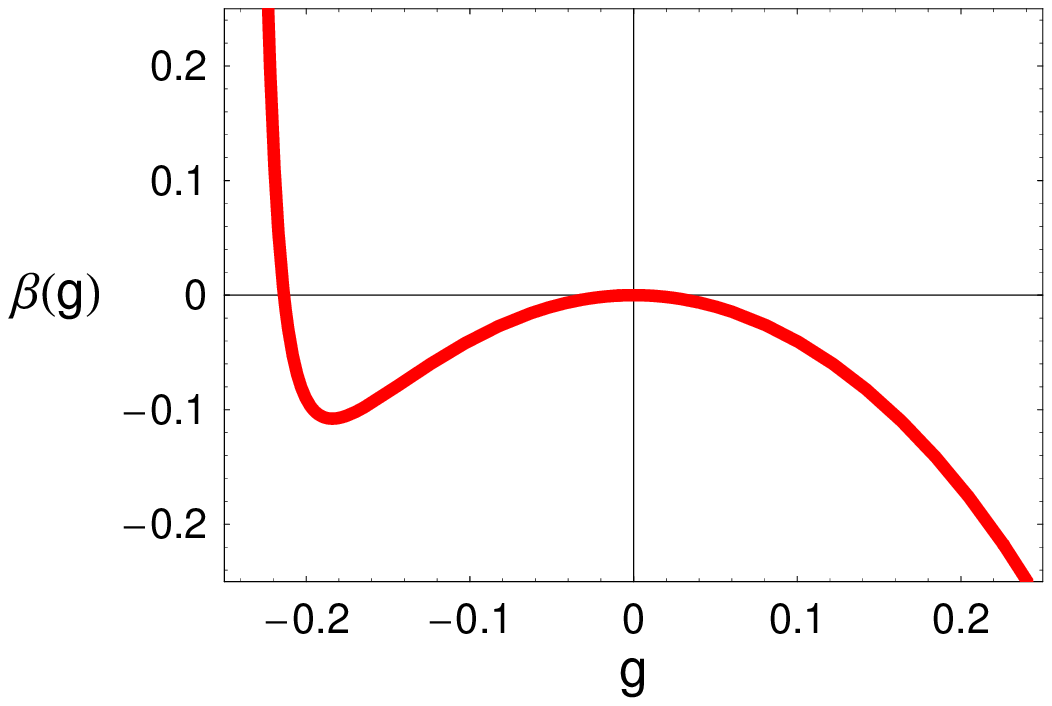}}\ \\
  \centerline{{\large (A)} \hskip2.9in {\large (B)}}
  \caption{(A) $\beta(g)$ versus $g$ for the bosonic $AdS_5$ \NLsM\ ($D=4$).  The non-trivial zero is at $g \approx -0.198$.  (B) The analogous results for the type~II $AdS_5$ \NLsM.  The zero is at $g \approx -0.217$.}\label{BosonicSUSY}
 \end{figure}

The singular behavior of $b_2(\kappa)$ is crucial to the question of whether there is a zero of $\beta(g)$ for negative $g$ in $d=2$.  In the expansion
 \eqn{BetaExpandAgain}{
  {\beta(g) \over g} = \sum_{n=0}^\infty
    {b_n(\kappa) \over D^n} \,,
 }
if one stops after the term $b_1(\kappa)/D$, then as one decreases
$g$ from $0$ to negative values, $\beta(g)$ becomes singular (at
$\kappa=-3$, as explained above) before it has a zero.  But if one
includes also the term $b_2(\kappa)/D^2$, then the competition
between positive $b_0(\kappa)$ and negative $b_2(\kappa)$ leads to a
zero before the singularity at $\kappa=-1$.  In short, the behavior
in figures~\ref{FixedOrderBeta} and~\ref{BosonicSUSY} is typical.

If one further expands
 \eqn{bnExpand}{
  b_n(\kappa) = \sum_{k=n+1}^\infty b_{n,k} \kappa^k
 }
then the singular behaviors \SeveralSeries\ translate into the statements
 \eqn{bnkAsymptotics}{
  b_{1,k} = P_1(k) (-1/3)^k \qquad b_{2,k} = P_2(k) (-1)^{k+1} \,,
 }
where $P_1(k)$ and $P_2(k)$ are positive for all but finitely many $k$ and have at most polynomial growth.  Physically, the zero of the beta function computed through order $D^{-2}$ arises not so much from competition of one-loop and two-loop terms as in \cite{Banks:1981nn}, but more so from competition of the one-loop result from the asymptotics of high loop orders---or, in spacetime language, from competition between the Einstein-Hilbert term and high powers of the curvature tensor.

The outstanding question, of course, is what happens at higher
orders in $1/D$.  We have no real answers in the bosonic case, and
only partial information in the supersymmetric case, but let us
contemplate the two main alternatives.
 \begin{enumerate}
  \item Most optimistic is the supposition that higher orders in $1/D$ are no more singular than $b_2(\kappa)$, so that the beta function is well-approximated by the $1/D^2$ result that we have developed in this section.  This would seem more plausible if it could be shown that general amplitudes contributing to $\lambda_i(\epsilon)$ have their singularities at integer $\epsilon$, or, equivalently, integer $d$.
  \item Alternatively, one could imagine that the $b_n(\kappa)$ become singular at progressively small values of $\kappa$---for example, at $\kappa = -3/(2n+1)$.  Then for finite $D$, $\beta(g)$ does not have a convergent power series expansion.  This would not be atypical of perturbive expansions.\label{BadCase}
 \end{enumerate}
Even in case (\ref{BadCase}), it could still be that correct qualitative information can be gleaned from a partial summation of the beta function, such as the $1/D^2$ results that we have in hand.  After all, fairly good agreement is obtained for the $O(N)$ model between these results in $d=3$ and other treatments \cite{Vone,Vtwo}.

Because $\beta(g)$ computed through $1/D^2$ slopes steeply down at its non-trivial zero, $g_c < 0$, the critical exponent $\lambda = -{1 \over 2} \beta'(g_c)$ is large and positive.  The corresponding operator then must have a dimension which is large and negative: see figure~\ref{BosonicSUSY} and table~\ref{CCTable}.  This appears to violate unitarity, and we might wonder anew whether the fixed point really exists.  Let us reflect, however, on another CFT with non-compact target space, namely the free massless boson $X$ in $d=2$.  The correlator $\langle X(x) X(0) \rangle \sim \log |x|$ grows at large $|x|$, signalling large infrared fluctuations.  $AdS_{D+1}$ is in some sense much ``bigger'' than ${\bf R}^{D+1}$, because the volume enclosed within a radius $\ell$ of a given point grows exponentially with $\ell$ rather than as a power.  So we should not be too surprised to find even wilder infrared fluctuations, mediated perhaps by operators whose two-point correlators do grow as positive powers.  Two-point functions that increase weakly with distance were found in \cite{Duncan:2004ik}, and their observation that the $AdS_{D+1}$ \NLsM\ has non-normalizable ground states appears to dovetail with the expectation of large infrared fluctuations.

We are also struck by the observation that the integral form
\IntegratedBs\ for the beta function coefficients must draw a large
contribution from regions where $\epsilon$ is close to $\kappa_c$,
which is finite and negative---in fact, $\kappa_c$ is close to $-1$
when $D$ is very large.  Perhaps this means that configurations in
target space with a Hausdorff dimension closer to $1$ than $2$ make
a large contribution to the path integral.  Such configurations
would be somewhere between smooth surfaces (dimension $2$) and
branched polymer configurations (dimension $1$).  It is natural to
expect positive power laws in correlators for field theories in
$d<2$ dimensions.  Perhaps strong infrared fluctuations in the
$AdS_{D+1}$ target space result in an effective lowering of the
dimension in which the field theory is defined.

\subsubsection{Central charge of the non-trivial fixed point}
\label{CENTRALCHARGE}

Besides dimensions of operators at the fixed point, another
scheme-independent quantity is central charge.  The non-trivial fixed point is UV stable, and if it is perturbed slightly toward smaller $g$, there is an RG flow to flat space.  Because the beta function is roughly speaking the gradient of the central charge, the non-trivial fixed point has a higher central charge than flat space of the same dimension.  More precisely,
 \eqn{cBeta}{
  {\partial c \over \partial g} &= {3(D+1) \over 2g^2} \beta(g) \,.
 }
We defer a derivation of \cBeta\ until after \eno{cBoundBosonic}.  The prefactor comes from the metric on the space of couplings, and it may be corrected by loop effects.  As we will now see, leading order expressions are sufficient to make a reasonable estimate of $c$ at the non-trivial fixed point, at least for $D$ large.  Indeed, integrating across the RG flow from the non-trivial fixed point to flat space leads to
 \eqn{FindCC}{
  c &= D+1 + {3(D+1) \over 2} \int_0^{\kappa_c}
     {d\kappa \over \kappa} {\beta(g) \over g}  \cr
    &= D+1 + {3(D+1) \over 2} \int_0^{\kappa_c}
     {d\kappa \over \kappa} \left[ -\kappa +
      {1 \over D} b_1(\kappa) + {1 \over D^2} b_2(\kappa) \right]
        \cr
    &\approx (D+1) \left[ 1 - {3\kappa_c \over 2} \right]
 }
where in the second equality we have used \betaExpress.  In the approximate equality we simply note that the integrand is nearly constant over nearly the entire range of integration: it is dominated by the one-loop term in $\beta(g)$.  The main way in which the higher loop terms participate is in fixing $\kappa_c$.  In table~\ref{CCTable}, we have obtained more precise results by numerically integrating the second line of \FindCC.  Evidently,
 \eqn{cBoundBosonic}{
  c \leq {5 \over 2} (D+1) \,,
 }
and this bound saturates in the limit of large $D$.  It seems to us likely that $c$ does not exceed the the bound \cBoundBosonic\ even when higher $1/D$ corrections and a more precise treatment of ${\cal G}^{ij,kl}$ are included, for two reasons: first, such corrections cannot move $\kappa_c$ to a value less than $-1$; and second, they have little chance of making the integrand in \FindCC\ significantly larger over an appreciable range.

Let us now return to the derivation of \cBeta.  A general metric perturbation of flat space is accomplished through
 \eqn{VaryS}{
  \delta S = \int d^2 z \, \delta G_{ij} \, {\cal O}^{ij}
   \qquad
  {\cal O}^{ij} = {1 \over 2\pi\alpha'}
   \partial X^i \bar\partial X^j \,,
 }
where we employ the normalizations of \cite{PolchVolOne}.  As explained, for example, in section 15.8 of \cite{PolchVolTwo},
 \eqn{cGradient}{
  {\partial c \over \partial G_{ij}} &=
    24 \pi^2 {\cal G}^{ij,kl} \beta_{kl} =
    {3 \over 2} \beta^{ij}  \cr
  {\cal G}^{ij,kl} &= |z|^4 \langle {\cal O}^{ij}(z,\bar{z})
     {\cal O}^{kl}(0,0) \rangle =
    {\delta^{ik} \delta^{jl} \over 16\pi^2} \,,
 }
where in the second step of the first line we have used the final expression in the second line for the metric on the space of couplings, ${\cal G}^{ij,kl}$.  Note that ${\cal G}^{ij,kl}$ is computed in free field theory: this is why we remarked below \cBeta\ that it may suffer loop corrections.

Let us now apply the result ${\partial c \over \partial G_{ij}} = {3 \over 2} \beta^{ij}$ to the $AdS_{D+1}$ \NLsM.  We express the metric as $G_{ij} = -{1 \over g} G^{(0)}_{ij}$, where $G^{(0)}_{ij}$ is the metric on $AdS_{D+1}$ with radius of curvature $\sqrt{\alpha'}$ rather than $L$.  Thus $G_{ij} \partial / \partial G_{ij} = -g \partial/\partial g$, and we find
 \eqn{GettingPF}{
  g {\partial c \over \partial g} =
    -{3 \over 2} G_{ij} \beta^{ij} =
    {3 \over 2} {D+1 \over g} \beta(g) \,,
 }
where in the last step we used \ExpressBetaBeta, which is an exact expression.  A non-trivial check on the calculation is to compare a series expansion of $c$ in small $g$,
 \eqn{cSeries}{
  c = (D+1) - {3 \over 2} D(D+1) g - {3 \over 4} D(D+1) g^2 +
    \ldots
 }
with the tree-level spacetime effective action (see for example \cite{PolchVolOne})
 \eqn{SeffBosonic}{
  S = {1 \over 2\kappa^2} \int d^{D+1} x \, \sqrt{G}
    e^{-2\Phi} \left[ -{2 (D-25) \over 3\alpha'} + R +
     4(\partial\Phi)^2 + {\alpha' \over 4}
       R_{ijkl} R^{ijkl} + \ldots \right] \,.
 }
It is understood \cite{Jack1, Jack2, TseytlinCC} that, to the order
shown, in the scheme of minimal subtraction, and up to an overall
multiplicative factor, the quantity in square brackets in
\SeffBosonic\ must coincide with the central charge \cSeries, plus
$-26$ to account for reparametrization ghosts.  Our central charge
expression \cSeries\ indeed satisfies this constraint.  Furthermore,
any corrections to ${\cal G}^{ij,kl}$ in \cGradient\ must be at
least $O(\alpha'^2)$ for this matching to be satisfied.  Note that
this is an off-shell test: the comparison between \cSeries\ and
\SeffBosonic\ is being made for a very weakly curved $AdS_{D+1}$
space.
\begin{table}
\begin{center}
\begin{tabular}{|c||c|c|c||c|c|c|}
 \hline & \multicolumn{3}{|c||}{Bosonic} &
\multicolumn{3}{|c|}{Supersymmetric} \\ \hline $D+1$&$g_c
D$&$c$&$\lambda_c$&$g_c D$&$c$&$\lambda$\\ \hline
3 & -0.6603 & 5.228 & 1.698 & -0.8105 & 7.732 & 4.583 \\ \hline
4 & -0.7435 & 7.538 & 2.687 & -0.8447 & 10.58 & 6.026 \\ \hline
5 & -0.7926 & 9.899 & 3.734 & -0.8696 & {\bf 13.49} & 7.614 \\ \hline
6 & -0.8247 & 12.29 & 4.801 & -0.8872 & {\bf 16.42} & 9.209 \\ \hline
7 & -0.8473 & 14.69 & 5.866 & -0.9001 & 19.36 & 10.78 \\ \hline
8 & -0.8642 & 17.11 & 6.921 & -0.9101 & 22.3 & 12.32 \\ \hline
9 & -0.8773 & 19.54 & 7.96 & -0.918 & 25.26 & 13.82 \\ \hline
10 & -0.8877 & 21.97 & 8.982 & -0.9244 & 28.21 & 15.28 \\ \hline
11 & -0.8962 & {\bf 24.41} & 9.986 & -0.9298 & 31.17 & 16.71 \\ \hline
12 & -0.9034 & {\bf 26.85} & 10.97 & -0.9343 & 34.13 & 18.11 \\ \hline
13 & -0.9094 & 29.3 & 11.94 & -0.9381 & 37.1 & 19.48 \\ \hline
\end{tabular}
\caption{Values of the central charge and critical
exponent $\lambda$ in the bosonic and supersymmetric case for
various choices of $D$.  Included in the central charge in the
supersymmetric case is the fermionic contribution of $(D+1)/2$, as
well as some known $1/D^3$ contributions---see (\ref{betaApprox}).
We compute $\lambda$ as $-\beta'(g_c)/2$.  In bold we show the cross-over points between sub-critical and super-critical values of the central charge for both the bosonic and supersymmetric cases.}\label{CCTable}
\end{center}
\end{table}

\subsection{Explicit results for the $(1,1)$ supersymmetric case}
\label{SUSY}

Calculations analogous to those in sections~\ref{LOWESTORDER} and
\ref{NEXTORDER} were carried out in a series of papers by Gracey
\cite{gracey-short, gracey-long, gracey-eta, gracey-nu}.  The
computations were done for the $(1,1)$ supersymmetric $O(N)$ \NLsM\
in component formalism, starting with the action \OneOneComponents.
The final results for $\lambda$ are:
 \eqn{TypeIILambdas}{
  \lambda_0 &= \mu-1  \cr
  \lambda_1 &= 0  \cr
  \lambda_2 &=
   \frac{8 (\mu -1) \Gamma (2 \mu -2)^2}{\Gamma (2-\mu )^2 \Gamma (\mu -1)^4 \Gamma (\mu )^2} \Bigg[
   -\frac{4 \left(\psi(2-\mu )-\psi(\mu )+\psi(2 \mu -1)-\psi(1) \right)}{\mu -1}
    \cr &\qquad{}
   -2 \left(\psi(2-\mu )-\psi(\mu )+\psi(2 \mu -1)-\psi(1) \right)^2
    \cr &\qquad{}
   +2 \psi'(2-\mu )+5 \psi'(\mu )-2 \psi'(2 \mu -1)-5\psi'(1)
    \Bigg] \,.
 }
The relative simplicity of \TypeIILambdas\ over \AnomCoefs\ and \GotDeltasSecond\ results from non-trivial cancellations among several dozen Feynman graphs.  Because $\lambda_1=0$, we have
 \eqn{TypeIIBs}{
  {\beta(g) \over g} &= \epsilon-\kappa +
   {1 \over D^2} b_2(\kappa) + O(D^{-3})  \cr
  b_2(\kappa) &= -2\kappa \int_0^\kappa d\xi
    {\lambda_2(\xi) \over \xi^2} \,.
 }
It is interesting to examine the structure of the series expansion of $\beta(g)$ computed through order $1/D^2$: from \TypeIILambdas\ and \TypeIIBs,
 \eqn{BetaSeries}{
  \beta(g) &= - D g^2 - \frac{3}{2} \zeta(3) D^2 g^5 +
     \frac{27}{32}\zeta(4) D^3 g^6 -
     \frac{29}{20}\zeta(5) D^4 g^7  \cr
    &\qquad{}+ \frac{1}{192}\left[245 \zeta(6) +
       104 \zeta(3)^2\right] D^5 g^8 -
     \frac{1}{224}\left[311 \zeta(7) +
      156\zeta(3)\zeta(4)\right]D^6 g^9 + \ldots \,.
 }
Just as in section~\ref{CONSISTENCY}, one can make a non-trivial
check of \TypeIILambdas\ and \TypeIIBs\ by comparing \BetaSeries\
with the third line of \SeveralBetas.  The four-loop term---related
to the famous $\alpha'^3 \zeta(3) R^4$ term in the type~II string
theory action---agrees up to a term of order $D g^5$, which
corresponds to an $O(D^{-3})$ contribution to
$\lambda$.\footnote{Recall that in the dimensional regularization
scheme being used, it is the Riemann tensor, rather than the Weyl
tensor, that appears in this $R^4$ term.}  It is evident from
\BetaSeries---and it can be proven starting from
\TypeIILambdas---that the coefficient of the term $D^{k-2} g^{k+1}$
in \BetaSeries\ is a polynomial in the transcendental numbers
$\zeta(q)$ for integers $q>2$, such that each term is a rational
multiple of $\zeta(q_1) \zeta(q_2) \cdots \zeta(q_s)$ with $\sum_r
q_r = k-1$.\footnote{To see this, note that from \TypeIIBs\ that the
$g^{m+2}$ term of the beta function comes from the $\epsilon^m$ term
in the series expansion of $\lambda_2$ around $\mu=1+\epsilon/2=1$.
To prove the claim regarding the $\zeta$ dependence, start by noting
that the series expansion of the term in square brackets in
\TypeIILambdas\ can be written as
 \eqn{SquareBrackets}{
  \left[\ldots\right] = \sum_{k=1}^\infty a_k \epsilon^k \zeta(k+2) + \sum_{k=4}^\infty
    \sum_{k'=3}^{k-1} b_{k k'} \epsilon^k \zeta(k')\zeta(k-k'+2) \,,
 }
where the $a_k$ and $b_{k k'}$ are rational.  Next, rewrite the remaining factors as
 \eqn{RemainingFactors}{
  \frac{\epsilon^3}{4}\frac{\Gamma(1+\epsilon)^2}{\Gamma(1-\epsilon/2)^2\Gamma(1+\epsilon/2)^6} \equiv \frac{\epsilon^3}{4} F(\epsilon) \,.
 }
Note that $F(0)=1$.  Furthermore, $F'(\epsilon) = F(\epsilon)
\left[2\psi(1+\epsilon) + \psi(1-\epsilon/2) -
3\psi(1+\epsilon/2)\right] \equiv F(\epsilon) G(\epsilon)$. Next,
note that $G(0)=0$, $G'(0)=0$, and $G^{(n)}(0) \propto \zeta(n+1)$
with non-zero, rational proportionality, since $\psi^{(n)}(1) =
(-1)^{n+1} n! \zeta(n+1)$.  Finally, $F^{(n)}(0)$ is equal to
$F(0)=1$ times a polynomial of derivatives of $G$ at $\epsilon=0$,
such that the number of derivatives plus factors of $G$ in a given
term sums to $n$.  These factors can then be easily recompiled to
prove the claim.}  The significance of this will emerge in
section~\ref{STRINGLOOPS}.

It is also evident from \BetaSeries\ that after the Einstein term
$-Dg^2$, the sign of each term alternates, in such a way that for $g
< 0$,
 \eqn{BetaSimpleSeries}{
  \beta(g) = -c_1 |g|^2 + c_4 |g|^5 + c_5 |g|^6 + c_6 |g|^7
    + \ldots \,,
 }
where $c_k > 0$ for all $k$.  Thus there must be a zero of the beta function (to the order we have computed it) before there is a singularity: the first term balances against all the rest.  This pattern of signs is related to the fact that $b_2(\kappa)$ is positive and singular at $\kappa=-1$:
 \eqn{bTwoSingular}{
  \lambda_2(\epsilon) = {4/\pi^4 \over (\epsilon+1)^4} +
    O\big[ (\epsilon+1)^{-3} \big] \qquad
  b_2(\kappa) = -{8/3\pi^4 \over (\kappa+1)^3} +
    O\big[ (\kappa+1)^{-2} \big] \,.
 }
As in the bosonic case, $b_2(\kappa)$ has all its singularities on the real axis, and $\kappa=-1$ is the one closest to the origin.

The distinctive singular behavior at $\kappa=-1$ is the same as in the bosonic case: see figure~\ref{BosonicSUSY} and table~\ref{CCTable}.  And, just as in the bosonic case, we must be careful to
qualify the claim that there is a zero of the beta function with the
caution that higher order terms in the $1/D$ expansion could change
the story.  The particular danger emphasized in
section~\ref{SINGULARITY} was that $b_3(\kappa)$ and/or higher
$b_n(\kappa)$'s could become singular for less negative values of
$\kappa$ than $-1$.  In the all-orders beta function, expressed as a power series as in \BetaSimpleSeries, each coefficient $c_k$ is a polynomial in $D$.  The large $D$ calculations performed to date (i.e.~through $O(D^{-2})$) tell us only the leading behavior of each $c_k$ for large $D$.  Terms in these polynomials which are subleading in $D$ could nevertheless eventually dominate the contribution of high loop terms and control the existence of a fixed point.  The two main alternatives contemplated in section~\ref{SINGULARITY} are alternatives still for the supersymmetric case.

In an attempt to obtain partial information available about
$b_3(\kappa)$ and other high order terms, Gracey \cite{gracey-short}
has noted that consistency with the results of \cite{Morozov:1984ad}
demands that the tensor structure $R^{(n)}_{ab}$ entering into
$\beta(g)$ for a general homogenous space at $k=n+3$ loops ($n>0$) should vanish on K\"ahler manifolds.  He noted that a satisfactory form of
the $R^{(n)}_{ab}$ is
 \eqn{HigherTensors}{
  R_{a b}^{(n)} =
   R_{acde}R_{bpq}{}^e \left[ R^c{}_{l_1 m_1}{}^p R^{q l_n m_n d}
      + R^c{}_{l_1 m_1}{}^q R^{dpm_n l_n} \right]
     \prod_{i=1}^{n-1} R_{l_{i+1}}{}^{l_i m_i}{}_{m_{i+1}} \,.
 }
If the $R^{(n)}_{ab}$ were the only tensor structures contributing to the beta function at loop order $k=n+3$, then the coefficients $c_k$ would not change sign as compared to their behavior at leading order in $D$: indeed,
 \begin{eqnarray*}
  \mbox{$n$ odd:  }& R_{a b}^{(n)} = & (g^{n+3} G_{a b}) \left[2(D-1)
     D^3 \sum_{k=0}^{(n-3)/2} D^{2k} + 3D(D-1)\right]\\
  \mbox{$n$ even:  }& R_{a b}^{(n)} = & -(g^{n+3} G_{a b})
    \left[ 2 (D-1) D^2 \sum_{k=0}^{(n-2)/2} D^{2k} \right] \,.
 \end{eqnarray*}
In terms of the large $D$ expansion, we may summarize the discussion as follows:
 \eqn{betaApprox}{
  {\beta(g) \over g} \approx \epsilon-\kappa + {D-1 \over D^3}
    b_2(\kappa) \,.
 }
It is not claimed that the form \betaApprox\ entirely accounts for $O(D^{-3})$ effects, but it does capture the four-loop term in \SeveralBetas\ exactly, and it also correctly captures the one-loop exactness of the $O(3)$ model \cite{Novikov:1983mt}.  And, of course, it retains the zero for negative $\kappa$ which has been our main interest.

There is another constraint on tensor structures that contribute to
the beta function at higher loops in a minimal subtraction scheme:
they cannot involve factors of the Ricci scalar or Ricci tensor.  We
learn this from the background field method \cite{Jack1,
alvarez-gaume}.  In this scheme, the beta function depends only on
the simple poles of the dimensional regularization parameter
$\epsilon$ in Feynman diagrams that have only ``external vertices".
Each external vertex comes with one or two powers of the Riemann
tensor. Factors of the Ricci scalar or tensor would come from lines
that loop back to the same vertex---propagators of internal lines
proportional to $\delta_{ij}$ result in self-contractions of the
Riemann tensor.  Such a loop contributes a simple pole in
$\epsilon$. The remainder of the diagram must contribute at least a
simple pole in $\epsilon$, hence the entire diagram has a pole of at
least $\epsilon^{-2}$, and therefore does not contribute to the beta
function.

As in the bosonic case (see section~\ref{SINGULARITY}) the operator controlling the leading corrections to short-distance scaling is predicted to have negative dimension.  We offer again the suggested interpretation that large infrared fluctuations, possibly leading to an embedding in target space with Hausdorff dimension less than $2$, lead to the power-law growth in certain two-point functions that negative operator dimensions imply.

The central charge computation also goes through almost as in section~\ref{CENTRALCHARGE}.  The main difference is that the $\kappa\to 0$ limit of the central charge is ${3 \over 2} (D+1)$ rather than $D+1$ due to the contribution of fermions.  A free-field treatment of the metric on the space of couplings leads to
 \eqn{cBetaII}{
  {\partial c \over \partial g} &= {3(D+1) \over 2g^2} \beta(g) \,,
 }
identically to the bosonic case \cBeta.  We defer a justification of \cBetaII\ until after \eno{cBoundII}.  Integrating \cBetaII,
 \eqn{FindSUSYCC}{
  c &= {3 \over 2} (D+1) + {3(D+1) \over 2} \int_0^{\kappa_c}
     {d\kappa \over \kappa} {\beta(g) \over g}  \cr
    &= {3 \over 2} (D+1) + {3(D+1) \over 2} \int_0^{\kappa_c}
     {d\kappa \over \kappa} \left[ -\kappa +
      {D-1 \over D^3} b_2(\kappa) \right]
        \cr
    &\approx {3 \over 2} (D+1)
      \left[ 1 - \kappa_c \right] \,,
 }
and we conclude that
 \eqn{cBoundII}{
  c \leq 3 (D+1) \,,
 }
with the bound saturating as $D \to \infty$.  In table~\ref{CCTable}, we have obtained more precise results by numerically integrating the second line of \FindSUSYCC.

Let us now return to the derivation of \cBetaII.  The considerations of \VaryS\ and \cGradient\ generalize straightforwardly: the metric perturbation operator is
 \eqn{MetricPerturbationII}{
  {\cal O}^{ij} = {1 \over 4\pi} \left( {2 \over \alpha'}
    \partial X^i \bar\partial X^j + \psi^i \bar\partial \psi^j + 
    \tilde\psi^i \partial \tilde\psi^j \right)
 }
The fermion terms contribute only contact terms to $\langle {\cal O}^{ij}(z,\bar{z}) {\cal O}^{kl}(0,0) \rangle$, so \cGradient is unchanged.  The discussion leading to \GettingPF\ is also unchanged.  Series expansions show that
 \eqn{cSeriesII}{
  c = {3 \over 2} (D+1) - {3 \over 2} D(D+1) g -
   {9 \over 16} \zeta(3) D(D^2-1) g^4 + O(g^5) \,.
 }
To match this onto a spacetime action for non-critical type~II superstrings, we require the form
 \eqn{SII}{
  S = {1 \over 2\kappa_D^2} \int d^D x \, \sqrt{G} e^{-2\Phi}
   \left[ -{2(c+c_{\rm ghost}) \over 3\alpha'} + R + 
    4(\partial\Phi)^2 + \ldots \right] \,.
 }
Then, up to an overall factor and the ghost contribution, the first two terms of \cSeriesII\ match the first two terms in square brackets of \SII.  We note that \SII\ is of exactly the same form as \SeffBosonic\ to the order shown.

There seem to be some conflicting claims in the literature\footnote{For instance, in \cite{Silverstein:2001xn, Maloney:2002rr}, a potential is quoted which is $2/3$ of the one we claim, and in \cite{deAlwis:1988pr}, the result is $1/2$ of the one we claim.  But we find consistency with the results of \cite{Myers:1987fv}.} about the normalization of the potential term for non-critical superstrings.  This normalization is tied to the factor in \cBetaII, so it is crucial for our discussion of central charges to get it right.  Let us therefore consider an alternative to the straightforward free-field worldsheet calculation discussed after \MetricPerturbationII.  Namely, in the NS5-brane background of type~II string theory, $c = 9$ comes from the ${\bf R}^{5,1}$ portion of the CFT; $c = 9/2 - 6/k$ comes from the supersymmetrized WZW theory describing the $S^3$ with $k$ units of $H_3$ flux (so that there are $k$ coincident NS5-branes); and $c = 3/2 + 6/k$ is supposed to come from a linear dilaton background, for a total of $c=15$ (see for example \cite{Callan:1991at}).  The geometrical description of the throat geometry is
 \eqn{ThroatDilaton}{
  ds^2 = -dt^2 + d\vec{x}^2 + dr^2 + k\alpha' d\Omega_3^2 \qquad
   \phi = -r/\sqrt{k\alpha'} \qquad H_3 = -k\alpha' \vol_{S^3} \,,
 }
where $\vec{x}$ is a vector in ${\bf R}^5$.  Consider now a spacetime treatment of the radial direction only.  The string frame action (consistent with \SII) is
 \eqn{RadialAction}{
  S = \int dr \, e^{-2\Phi} \left[ -{2\delta c \over 3\alpha'} + 
   4(\partial\Phi)^2 \right] \,,
 }
where $\delta c = -6/k$ is the central charge deficit from the supersymmetrized WZW theory that forces a dilaton flow upon us.  The radial equation of motion for the dilaton derived from \RadialAction\ indeed possesses the solution $\phi = -r/\sqrt{k\alpha'}$, providing a check on the normalization of the potential term.

\section{Discussion}
\label{DISCUSSION}

In sections~\ref{FIXEDORDER} and~\ref{LARGED}, we have laid out the evidence that $AdS_{D+1}$ \NLsM's, both supersymmetric and non-supersymmetric, have zeroes at finite coupling, at least for large enough $D$.  We have also explained how this evidence could be misleading: see in particular the discussion following \BosonicBeta\ and in section~\ref{SINGULARITY}.  While cautioning once more that we have not shown that our calculations rest upon a controlled approximation scheme, we will in this section assume what seems to us the likeliest alternative: that there are indeed non-trivial fixed points.  What then are the consequences? 

First and foremost, there may be vacua of string theory which incorporate an $AdS_{D+1}$ \NLsM\ in the worldvolume theory.  In section~\ref{YANGMILLS} we focus on possible implications of an $AdS_5$ vacuum, and in section~\ref{QUOTIENTS} we turn to vacua with an $AdS_3$ factor.  Aspects of these discussions will be quite speculative.  In section~\ref{POSITIVE} we comment on a sort of no-go theorem for de Sitter vacua arising from competition between different powers of the curvature.  And in section~\ref{STRINGLOOPS} we remark on a prescription for including the effects of string loops in the spacetime effective action.

\subsection{A dual for pure Yang-Mills theory?}
\label{YANGMILLS}

The many existing examples of anti-de Sitter string vacua involve matter fields whose stress-energy supports the negatively curved geometry.  In terms of the \NLsM, there are additional couplings of the scalar fields (and possibly the fermions in the supersymmetric case) that modify the one-loop beta function in such a way that there is fixed point---usually, in fact, a fixed line.  Actually, this is an over-simplification in most cases: the matter fields are usually Ramond-Ramond fields, and including them in the action leads to technical difficulties; see for example \cite{Berkovits:1999im}.  In any case, a string vacuum based on an $AdS_{D+1}$ \NLsM\ of the unadorned type that we have described would be quite different from most previously studied cases, for example in that the scale of curvature is necessarily close to the string scale.  More precisely, $|\kappa| = \alpha' D/L^2 \approx 1$, so the radius of curvature of $AdS_{D+1}$ is $L \approx \sqrt{\alpha' D}$.  Because the construction we have given makes no reference to D-branes, it is not obvious what the dual field theories in $D$ dimensions should be.  In this respect our story is similar to \cite{Polyakov:2004br}; see also \cite{Polyakov:2005ss} for more recent developments.  

But a two-dimensional worldsheet CFT is not enough to guarantee the existence of a string vacuum.  One must also cancel the Weyl anomaly and define physical state conditions in a consistent way.  In general, one must impose a GSO projection that leads to a closed OPE, modular invariance, and, if stability is required, no relevant deformations of the worldsheet theory.

The results for the central charge listed in table~\ref{CCTable} are not particularly encouraging at first sight: $c$ is always non-integer.  But recall that a spatial dilaton gradient continuously increases the central charge.  This is best understood in free field theory (the linear dilaton vacuum) but based on the spacetime effective action, it seems obviously to remain true in curved backgrounds.  As long as we focus on Euclidean theories, decreasing the central charge with a timelike dilaton gradient is not an option.  Referring to table~\ref{CCTable}, we see that we can cancel the Weyl anomaly in the type~II case with a dilaton gradient for $AdS_3$, $AdS_4$, and $AdS_5$, but not $AdS_6$ and higher.  A natural choice is for the dilaton gradient to be chosen in the radial direction in a Poincar\'e patch description of the geometry.  The resulting geometry is an obvious candidate for a holographic dual of a gauge theory undergoing renormalization group flow.  It is pleasing to see this construction work in the dimensions that it does because it is in dimensions $4$ and lower that gauge theory interactions are renormalizable.

But wouldn't it be lovely if the Weyl anomaly canceled exactly in $AdS_5$, without any dilaton gradient?  That would mean that there is no potential for the dilaton.  Turning on a dilaton gradient would still be allowed, just as in $AdS_5 \times S^5$ \cite{Gubser:1999pk}.  Presumably, it would be a gentler flow, at least when $\Phi$ is large and negative.  This would correspond to the fact that gauge couplings in four dimensions experience logarithmic flow, which is qualitatively gentler than the power law flow in lower dimensions.

As things stand in table~\ref{CCTable}, the critical dimension for the string theory construction seems to be about halfway between $D=4$ and $D=5$ (a precise computation gives $D=4.517$).  Everything would fit together a bit better if the critical dimension for the string theory construction coincided precisely with the upper critical dimension for gauge interactions in the AdS/CFT duals.  If only \cBoundII\ were an equality rather than an inequality!  Recall that in deriving \cBoundII, we combined rather precise knowledge of $\beta(g)$ with a less precise computation of the metric on the space of couplings.  Perhaps then there is enough wiggle room to wind up with $c=3(D+1)$ for the non-trivial fixed point.  Or perhaps we are engaging in wishful thinking: after all, our calculations were accurate enough so that the series expansion \cSeriesII\ avoided $O(g^2)$ and $O(g^3)$ terms, which was necessary to match onto the absence of $R^2$ and $R^3$ terms in the type~II effective action.

It is particularly attractive to suppose that the critical background is $AdS_5$---probably with a slight dilaton gradient---rather than a product of $AdS_5$ and some other geometry.  The dual field theory may then be expected to be comparably simple, without (for example) any flavor symmetries.  Perhaps a definite understanding of what theory it is would come from studying the dynamics of open strings on some version of D3-branes appropriate to the $AdS_5$ \NLsM.  Absent such a discussion, let us assume that it is pure Yang-Mills theory, with coupling $g_{YM} = e^\Phi$.  Because of our limited understanding of the spacetime effective action and the lack of a D-brane construction, we cannot presently determine the central charge or the gauge group.  But because the geometry is at string scale and the dilaton can flow to large negative values, we may hope that it embraces the asymptotic freedom and confinement of pure glue in a single holographic description.  For example, confinement of external electric charges and screening of magnetic ones might be accounted for in terms of the radial variation of the tensions (as measured in Einstein frame) of fundamental strings and D1-branes, as in \cite{Gubser:1999pk}.

Before getting carried away with the idea that the type~II $AdS_5$ construction should be critical, let us note another possibility that fits much better with the current numerical results: $AdS_5 \times S^1$ may be a critical background.  The worldsheet boson parametrizing $S^1$ and its superpartner together contribute $c=3/2$, so using the value for $AdS_5$ from table~\ref{CCTable} leads to a total central charge of $14.99$.\footnote{It may seem striking that $AdS_5 \times S^1$ is the bosonic part of ${SU(2,2|2) \over SO(5) \times SU(2)}$, which was argued in \cite{Polyakov:2004br} to support a \NLsM.  But the reasoning in \cite{Polyakov:2004br} hinged on the inclusion of a fermionic version of a Wess-Zumino term whose presence combined with kappa symmetry forbade a non-zero beta function.  This is rather different from our analysis.  Moreover, ${SU(2,2|2) \over SO(5) \times SU(2)}$ was stated in \cite{Polyakov:2004br} to be a non-critical background.}  The dual four-dimensional theory should have a $U(1)$ symmetry.  Pure ${\cal N}=1$ Yang-Mills theory is an obvious candidate.  If it is indeed the dual theory, then a fuller treatment should reveal a dilaton gradient, a dynamical breaking of the $U(1)$ symmetry, and (most fundamentally) a GSO projection that leads to spacetime supersymmetry.  Indeed, an understanding of the GSO projection seems crucial to the entire discussion of $AdS_{D+1}$ string vacua, since we have to find some way to get rid of the negative dimension operator or otherwise ensure stability.

It is also somewhat striking that the cross-over from sub-critical to super-critical central charge happens for the bosonic string between $AdS_{11}$ and $AdS_{12}$.  Is it just coincidence that $11$ is also the preferred dimension of M-theory?

\subsection{Model building with quotients of $AdS_3$}
\label{QUOTIENTS}

Having articulated the hope that $c = 3(D+1)$ in a full calculation of the type~II $AdS_{D+1}$ \NLsM's, let us note another striking consequence of this conjectured equality: $AdS_3$ has $c=9$, the same value as a six-dimensional Calabi-Yau construction ($CY_3$).  We are thus led to suggest a new class of type~II string vacua: ${\bf R}^{3,1} \times AdS_3/\Gamma$, where $\Gamma \subset SO(3,1)$ is a discrete group such that $AdS_3 / \Gamma$ has finite volume.

There is an obvious candidate for a heterotic generalization, inspired by the familiar story of the standard embedding \cite{Candelas:1985en}.\footnote{We thank G.~Michalogiorgakis for a discussion on this point.}  Let the fields $X^\mu(z,\bar{z})$ and $\tilde\psi^\mu(\bar{z})$ form the usual free ${\bf R}^{3,1}$ portion of the theory, with $c=4$ and $\tilde{c}=6$.  Let the $AdS_3$ \NLsM\ be constructed in a symmetric fashion, so that $c=\tilde{c}=9$ (we assume).  The anti-holomorphic part of the Weyl anomaly is now canceled.  To cancel the holomorphic part, an additional holomorphic CFT is required with $c=13$.  This is easily big enough for model-building: for example, a Kac-Moody current algebra based on $E_8 \times SO(10)$ at level $1$ has $c=13$.  But we have no definite proposal for how an appropriately chiral spectrum of fermions might emerge in such a construction, nor for how to make an appropriate GSO projection in such a way as to get rid of tachyons and/or negative dimension operators and be left with a phenomenologically attractive gauge group.  

The total size of $AdS_3/\Gamma$ is fixed in string units, once $\Gamma$ is chosen, because the non-trivial zero of $\beta(g)$ is isolated.  We must of course assume that GSO has gotten rid of operators of dimension less than $2$.  A very attractive feature of the finite volume spaces $AdS_3/\Gamma$ is that they have no massless shape moduli \cite{mostow}.  There is however a large discrete class of these manifolds, and their number grows rapidly with their volume: see for example \cite{snappea}.  It is known that finite volume $AdS_3/\Gamma$ can never have Killing spinors \cite{Kehagias:2000dg,nasri}, so perhaps it is impossible to get ${\cal N}=1$ supersymmetric vacua in four dimensions.

Another notable feature of manifolds $AdS_3/\Gamma$ is that their volume scales exponentially with their linear size.  This has been exploited in the context of brane-world scenarios \cite{TroddenOne,nasri} to give an account of the hierarchy between the four-dimensional Planck scale and the weak scale without extensive tuning the size of the extra dimensions.  We also note that the implications for compact hyperbolic manifolds in cosmology has been the subject of many studies \cite{kaloper, starkmanOne, starkmanTwo, Townsend:2003fx, Ohta:2003pu, Roy:2003nd, Emparan:2003gg, Chen:2003dc}, and considerations for particle physics were discussed in \cite{Demir:2001ap}.  And we note that we evade the objections of \cite{Carroll:2001ih} because our construction is near the string scale.  But we cannot claim to have demonstrated stability until a satisfactory GSO projection is formulated.

The framework we have suggested so far seems to depend strongly on having $c=9$ exactly for the type~II $AdS_3$ \NLsM.  This value is surely within ``theoretical errors'' of the results listed in table~\ref{CCTable}: large $D$ methods cannot be expected to be terribly accurate for $D=2$!\footnote{Indeed, we should admit the possibility that the fixed point exists for larger $D$, but not $D=2$.  Certainly it doesn't exist for $D=1$: see the discussion preceding \HigherTensors.}  But there might be some interest in this type of construction even if this is not true.  Perhaps some flux of $H_3$ can be used to adjust the central charge.  Perhaps, in a case where $AdS_3/\Gamma$ has finite volume but is not compact, a dilaton gradient might be added.  If the value $c=7.732$ is a slight {\it over-estimate} of the true value, then perhaps a compactification to five dimensions (possibly with a slight dilaton gradient) would be possible.

In any case, the large discrete freedom in the choice of $AdS_3/\Gamma$ could result in an interesting range of four-dimensional models, as in \cite{Bousso:2000xa}.  Most finite volume hyperbolic manifolds are not simply connected,\footnote{One particularly symmetrical example has homotopy group ${\bf Z}_3 + {\bf Z}_3 + {\bf Z}_3$ \cite{snappea}.} and the minimal length of the various homotopy cycles are fixed in string units.  This suggests interesting possibilities for the spectrum of wrapped strings.  For example: could they provide generations of chiral fermions?

We should also consider the possibility that type~II on ${\bf R}^{3,1} \times AdS_4$ is a critical background.  Although this is almost as close to being realized in table~\ref{CCTable} as criticality of ${\bf R}^{3,1} \times AdS_3$, it doesn't fit with the attractive hypothesis that $AdS_5$ is critical (which works if the values for $c$ in table~\ref{CCTable} are under-estimates); nor does it fit particularly well with the supposition that $AdS_5 \times S^1$ is critical, since this hypothesis works nearly perfectly when the values in table~\ref{CCTable} are assumed to be highly accurate.  Clearly, it would be highly desirable to determine the central charges more precisely.  Absent this, a more quantitative estimate of the theoretical uncertainties in the determinations in table~\ref{CCTable} would help bring the current discussion into better focus.

\subsection{Positive curvature spacetimes}
\label{POSITIVE}

It is well-accepted that there cannot be a conformal \NLsM\ on a sphere $S^{N-1}$ in $d=2$ without some participation of matter fields (i.e.~couplings other than the target space metric): such a \NLsM\ would be associated with a finite-temperature phase transition for the $O(N)$ model, which the Coleman-Mermin-Wagner theorem prohibits \cite{coleman,merminwagner}.  This theorem states (more or less) that a compact, continuous symmetry cannot exhibit spontaneous symmetry breaking in two dimensions.\footnote{Reasons are noted in \cite{Duncan:2004ik} why the CMW theorem does not extend to the non-compact symmetry group $SO(D+1,1)$ acting on $AdS_{D+1}$.}

It is only a small step from this to claim that there cannot be a conformal \NLsM\ on de Sitter space, $dS_{D+1}$, without the participation of some matter fields.  The reason, simply, is that the beta function is indifferent to signature.  If competition between different powers of curvature gave rise to a $dS_{D+1}$ fixed point, there would be a $S^{N-1}$ fixed point too (with $N=D+2$ as always).  Previous claims that de Sitter vacua {\it do} arise in this way \cite{Bento:1998bk} rely upon a $SL(2,{\bf Z})$-covariantization of the type~IIB spacetime effective action.  This takes us outside our current scope (see however section~\ref{STRINGLOOPS}), but we believe a careful analysis of signs still indicates $AdS_{D+1}$ rather than $dS_{D+1}$ solutions.

It is possible to check that the expressions \betaExpress\ and \betaApprox\ that we have given for the bosonic and type~II \NLsM's are consistent with the expectation that there are no positive curvature conformal points.  But there is a subtlety: for $d \geq 4$, the scalar field interactions of our model become irrelevant, and hence do not affect the renormalization group flow.  Thus for $d \geq 4$, we find the mean-field theory result $\lambda = 1$, to all orders in $1/D$, in both the bosonic and supersymmetric cases.  A check on the expressions \AnomCoefs, \GotDeltasSecond, and~\TypeIILambdas\ for the $\lambda_i$ is that $\lambda_0=1$ and $\lambda_1=\lambda_2=0$ for $d=4$, so that $\lambda = 1$.  In short, the large $D$ results match smoothly onto mean field theory expectations, order by order in $1/D$.  If this matching were not accounted for, then $\beta(g)$ would oscillate wildly for $\kappa > 2$, resulting in multiple zeroes.  When it is, $\beta(g)$ has no zeroes at all for $\kappa > 0$.

\subsection{Including String Loops}
\label{STRINGLOOPS}

Although our focus has been on non-trivial fixed points of $AdS_{D+1}$ \NLsM's, it is interesting to note that a series expansion of the  beta functions calculated in section~\ref{LARGED} can be translated into information about the low-energy effective action, or equations of motion, in ordinary type~II superstring theory in ten nearly flat dimensions.  We will focus on type~IIB and show how to arrange $SL(2,{\bf Z})$ invariance.

The tree-level effective action in string frame has the schematic form
 \eqn{StringActionTen}{
  S = {1 \over 2\kappa^2} \int d^{10} x \, \sqrt{G} e^{-2\Phi}
   \left[ R + a_4 R^4 + a_5 R^5 + \cdots \right] \,,
 }
neglecting terms which involve derivatives of curvature.  The coefficients $a_i$ are closely related to the coefficients in a power series expansion of $\beta(g)$ in $g$.  In Einstein frame, $g_{MN} = e^{-\Phi/2} G_{MN}$, and the term at order $R^k$ will pick up a dilaton dependence of $e^{(1-k)\Phi/2}$.  The Einstein-Hilbert term is then evidently $SL(2,{\bf Z})$-invariant, but the higher powers of curvatures are not.  An algorithm for rendering them $SL(2,{\bf Z})$-invariant has been discussed by several authors \cite{green-gutperle, Antoniadis:1997zt, Russo:1997fi}: briefly, one makes the replacement 
 \eqn{ZetaSub}{
  2\zeta(q)e^{-q\Phi/2} \rightarrow
     \sum_{(m, n) \neq (0,0)} \left|
       \frac{\sqrt{\mbox{Im } \tau}}{m + n \tau}\right|^q
   \rightarrow
     \sum_{(m, n) \neq (0,0)} \frac{e^{-q \Phi/2}}{(m^2 + n^2
       e^{-2\Phi})^{q/2}}
 }
in all the coefficients $a_k$.  The replacement \ZetaSub\ leads to an effective action with information at all orders in string loops.  For \ZetaSub\ to make sense, a highly non-trivial property is required of the $a_k$: they must be polynomials in $\zeta(q)$ such that every term is some multiple of a product $\zeta(q_1) \zeta(q_2) \cdots \zeta(q_r)$ such that $\sum_s q_s = k-1$.  Then, in Einstein frame, all the dependence of the coefficients of $R^k$ on the dilaton and on transcendental numbers can be factored into products of the form on the first expression in \ZetaSub.  The middle expression in \ZetaSub\ is $SL(2,{\bf Z})$-invariant: $\tau = C_0 + ie^{-\Phi}$ is the usual complexified dilaton.  In the last step, we specialize to configurations with $C_0=0$. 

As we noted in \BetaSeries\ and the discussion following it, the coefficient of $D^{k-2} g^{k+1}$ in $\beta(g)$ satisfies precisely the zeta function property required for \ZetaSub\ to make sense.  It is an interesting, qualitative, all-orders consistency check on the tree-level equation of motion $\beta(g)=0$ that one can straightforwardly render it $SL(2,{\bf Z})$-invariant.  

It is tempting to speculate that some version of $SL(2,{\bf Z})$-invariance can be arranged in $AdS_5$ (or perhaps $AdS_5 \times S^1$), where, according to our earlier speculations, the Weyl anomaly cancels just as in ten-dimensional flat space.  If so, then in the same spirit as \cite{Bento:1998bk}, it should be possible to solve the $SL(2,{\bf Z})$-covariant equation of motion for the complexified dilaton by setting $\tau=i$, i.e.~$\Phi=C_0=0$.  The reason is that $\tau=i$ is necessarily an extremum of an $SL(2,{\bf Z})$-invariant function.  The upshot is that by making the replacement \ZetaSub\ and then setting $\Phi=0$, one obtains from $\beta(g)=0$ a modified equation which still has an $AdS_5$ solution.  The radius is slightly larger: $\kappa_c \approx -0.72$ rather than the tree-level value $-0.8696$.

We are attracted to the idea that $SL(2,{\bf Z})$-invariance can be realized at some level for type~II strings on $AdS_5$, because in the conjectured dual, pure Yang-Mills theory, the action possesses an $SL(2,{\bf Z})$ invariance.  This does not imply that there is a conformal fixed point for this theory, but rather that renormalization group flows should have images under $SL(2,{\bf Z})$.  Whether the $AdS_5$ vacuum with $\Phi=0$ suggested in the previous paragraph is physically significant for pure Yang-Mills theory seems doubtful; more plausible is the idea that some dilaton gradient is required.  But we remind the reader that we are at the end of a long chain of conjectures.

\section{Conclusions}
\label{CONCLUSIONS}

The absence of a zero of $\beta(g)$ for positive $g$---that is, for the $O(N)$ model---is well accepted because it is consistent with the known perturbative results, the large $N$ results, the Coleman-Mermin-Wagner theorem, and, in the bosonic case, the non-perturbative results of \cite{Polyakov:1983tt}.  The situation for the $AdS_{D+1}$ \NLsM\ is much less certain.  The evidence from four-loop perturbative calculations and from the large $D$ expansion through order $1/D^2$ points to the following picture:
 \begin{enumerate}
  \item There is indeed a zero of $\beta(g)$ for negative $g$, of order $-1/D$.  This is true both for the bosonic case and the supersymmetric case, at least for large enough $D$.  In the supersymmetric case, $\beta(g)$ cannot have a zero for $D=1$ (i.e.~$AdS_2$), but for $D \geq 2$ it may.  See equations~\eno{IntegratedBs}, \eno{AnomCoefs}, \eno{GotDeltasSecond}, \eno{TypeIILambdas}, \eno{TypeIIBs}, and \eno{betaApprox}; figures~\ref{FixedOrderBeta} and~\ref{BosonicSUSY}; and table~\ref{CCTable}.
  \item The non-trivial fixed point has the peculiar property that the leading corrections to short-distance scaling are controlled by an operator of negative dimension.  In section~\ref{SINGULARITY} we have outlined a heuristic physical picture that could account for this peculiarity: it hinges on the idea that infrared fluctuations are quite wild in an $AdS_{D+1}$ target space, and that these large, non-Gaussian fluctuations make it possible for two point correlators to have power-law growth rather than power-law fall-off.
  \item The central charge of the non-trivial fixed point has $c \lsim {5 \over 2} (D+1)$ in the bosonic case, and $c \lsim 3 (D+1)$ in the supersymmetric case.  The dominant uncertainty in determining $c$ is from the metric on the space of couplings, which we have computed at the level of free field theory on the worldsheet rather than in a systematic expansion.  In the supersymmetric case, we have suggested that $c=3(D+1)$ exactly because it makes sense for the critical dimension for strings in $AdS_{D+1}$ to coincide with the upper critical dimension of gauge interactions in ${\bf R}^D$.  This conjectured equality has the striking consequence that $AdS_3$ and quotients of it have the same central charge as six-dimensional Calabi-Yau manifolds.
 \end{enumerate}
Of these claims, surely the oddest is the second.  But because the dimension of the operator is related to the slope of $\beta(g)$ at the fixed point, it is more sensitive to the nearly singular behavior of $\beta(g)$ there than the position $g_c$ of the fixed point is.  Thus one might hope that this ``problem'' goes away on its own.  The determination of the central charge is also less sensitive to the precise form of $\beta(g)$ near the fixed point than $\beta'(g_c)$.  If the values in table~\ref{CCTable} are fairly accurate, then $AdS_5 \times S^1$ rather than $AdS_5$ is the leading candidate for a critical type~II background.  As we have noted in section~\ref{YANGMILLS}, modifying our string-scale $AdS_{D+1}$ backgrounds with a dilaton gradient for $D \leq 4$ leads to appealing candidates for holographic duals of simple gauge theories, such as pure Yang-Mills theory, with no flavor symmetries.

We have been careful all along to point out the potential pitfalls of our calculations, but let us reiterate the main one: we are not calculating in a controlled approximation scheme.  This is well illustrated for fixed order computations in figure~\ref{FixedOrderBeta}, where successive terms alternate up to the fourth loop order.  The alternating signs problem is fixed in the large $D$ treatment: all powers of $g$ except the first few contribute with the same sign to $\beta(g)$ for $g<0$.  But then the issue becomes the radius of convergence in $\kappa = gD$ of higher $1/D$ corrections.  It would clearly be desirable to gain better control over these corrections.  Perhaps the class of graphs at $O(D^{-3})$ that contributes to the leading singular behavior of $b_3(\kappa)$ is small enough for explicit calculations to be tractable.  It is also clearly worthwhile to try to approach the same broad class of \NLsM's in different ways, such as algebraic methods, lattice simulations, or (for $g>0$) high temperature expansions.  Explorations of D-branes in these \NLsM's is clearly called for, as is an understanding of the possible GSO projections and their implications for stability.

Even the most conservative reading of our results strongly suggests that there is interesting structure in the beta function for the $AdS_{D+1}$ \NLsM\ which is not reliably visible at low loop orders.  While we admit the possibility that all the strange behavior---including singularities and an apparent zero---could be artifacts of the minimal subtraction scheme we chose, or of a zero radius of convergence for $\beta(g)$, we think it more likely that these features signal interesting physics.  A fixed point of RG is the simplest and most attractive possibility.  If this fixed point is where we think it is; if it leads to a dual at string scale to four-dimensional gauge theory at moderate coupling; if it provides an alternative route to four-dimensional models---then these \NLsM's and variants of them could become a central part of string theory.

\section*{Acknowledgements}

We would like to thank P.~Argyres, O.~DeWolfe, A.~Dymarsky, M.~Edalati,
I.~Klebanov, D.~Malyshev, L.~McAllister, G.~Michalogiorgakis, I.~Mitra, A.~Polyakov,
R.~Roiban, E.~Silverstein, H.~Verlinde, A.~Weltman, and B.~Zwiebel for useful
discussions. J.F.~would also like to thank the University of Cincinnati,
Columbia University, and organizers of TASI~2005 for their hospitality
while this work was in progress. 

This work was supported in part by the Department of Energy under Grant
No.\ DE-FG02-91ER40671, and by the Sloan Foundation.  The work of
J.F.~was also supported in part by the NSF Graduate Research Fellowship
Program.

\bibliographystyle{ssg}
\bibliography{extend}
\end{document}